\documentclass[twocolumn]{aastex631}

\usepackage{amsmath}
\usepackage{xcolor}
\usepackage{multirow}
\usepackage[normalem]{ulem}

\newcommand{\Ag}{A_{\mathrm{g}}}
\newcommand{\Rp}{R_{\mathrm{p}}}

\newcommand{\Fobs}{F_{\mathrm{obs}}}

\newcommand{\PhaseFn}{\Phi}
\newcommand{\rhosurf}{\rho_{\mathrm{surf}}}
\newcommand{\taufour}{\tau_{400}}
\newcommand{\Idisk}{I_{\mathrm{disk}}}
\newcommand{\DOLP}{\mathrm{DOLP}}
\newcommand{\phipl}{\phi_{\mathrm{pl}}}
\newcommand{\thetai}{\theta_{\mathrm{incl}}}
\newcommand{\vSmartMOM}{\texttt{vSmartMOM}}

\newcommand{\slopeI}{\mathcal{S}_I}
\newcommand{\slopeD}{\mathcal{S}_{D}}
\newcommand{\shapeI}{\mathcal{L}_{400}}
\newcommand{\obsvec}{\mathbf{m}}
\newcommand{\statevec}{\mathbf{x}}
\newcommand{\jac}{\mathsf{J}}

\begin{document}

\title{Multi-epoch ultraviolet observables for breaking the
       radius-albedo degeneracy in directly imaged exoplanets}

\author[0000-0003-0754-9154]{Suniti Sanghavi}
\affiliation{Jet Propulsion Laboratory, California Institute of
Technology, Pasadena, CA 91109, USA}

\author[0000-0002-4320-2599]{Robert A. West}
\affiliation{Formerly Jet Propulsion Laboratory, California Institute
of Technology, Pasadena, CA 91109, USA}

\author[0000-0003-1195-9666]{Pin Chen}
\affiliation{Jet Propulsion Laboratory, California Institute
of Technology, Pasadena, CA 91109, USA}

\begin{abstract}
{Direct imaging measures how bright a planet appears in reflected starlight, but brightness alone cannot tell whether the
planet is large and dark or small and bright. {This radius-albedo degeneracy limits the characterization
of non-transiting exoplanets, including those targeted by the
Habitable Worlds Observatory (HWO).} {From multi-epoch observations we
construct} ultraviolet observables that are independent of
planetary radius: {the normalized 400~nm lightcurve shape,
the ultraviolet intensity and polarization colors, and the degree of
linear polarization.} In a clear 360-400~nm spectral window
where Rayleigh scattering is strong and bright ultraviolet surfaces are
nearly colorless, these observables constrain the atmospheric column,
the surface reflectivity, and, when relevant, the observed phases before
radius is inferred. Using the GPU accelerated vector radiative
transfer model \vSmartMOM, we test
how uniquely these radius-free observables determine the scattering
state at signal-to-noise ratios (SNRs) of 5, 20, and 100. For
the chosen six phase sequence, spectropolarimetry gives median radius
consequences of 9.5\%, 3.3\%, and 0.45\% when the phases are known. {If the phases are unknown, six distinct epochs keep the
penalty modest at low SNR and negligible at moderate-high SNR, with
median consequences of 14\%, 3.2\%, and 0.47\%, respectively.} Removing polarimetry degrades the result,
especially in the HWO simulations where phase coverage is
restricted by inner working angle. For
solar twin systems at 6 and 12 parsecs, the required six-phase
campaigns fall in the allotted range of a few hundred hours, {making this a {clear} observing path to planetary radii
from HWO's reflected-light detections.}}
\end{abstract}


\section{Introduction}
\label{sec:intro}

The direct imaging of exoplanets in reflected starlight, the central
science driver of the Habitable Worlds Observatory \citep[HWO;][]%
{NAS2021decadal}, confronts a fundamental degeneracy. The reflected
flux measured at Earth from an unresolved planet scales as
\begin{equation}
\label{eq:fobs}
\Fobs(\lambda, \alpha) \;\propto\;
   \left(\frac{\Rp}{d}\right)^{\!2}
   \Ag(\lambda)\,\PhaseFn(\alpha),
\end{equation}
where $\Rp$ is the planetary radius, $d$ the star-planet separation,
$\Ag(\lambda)$ the geometric albedo, and $\PhaseFn(\alpha)$ the phase
function at phase angle $\alpha$. At known phase angle, or for an
adopted phase law, a single broadband measurement constrains only
the product $\Rp^2\,\Ag$, leaving radius and albedo entangled
\citep{Nayak2017, Feng2018}. {An unknown phase is an
additional quantity to infer in
Equation~\ref{eq:fobs}.} For a transiting planet the
radius can be determined from the transit light curve to break the degeneracy trivially, but the planets
HWO will image are overwhelmingly non-transiting, and their radii are
constrained only indirectly.

\subsection{State of the art}
\label{sec:intro:sota}

Reflected-light retrieval literature reveals various approaches to deal with the radius-albedo degeneracy: \citet{Lupu2016} avoided the full degeneracy by fixing the planetary radius, but still found strong correlations between cloud optical thickness and absorber abundances in reflected-light retrievals. \citet{Nayak2017} treated the radius as a free parameter along with the phase angle, finding that 
photometric observations constrain it only
to within a factor of two, with the residual ambiguity absorbed partially by the phase angle.
\citet{Feng2018} fixed the phase angle and freed the radius in a starshade rendezvous scenario for \textit{WFIRST}, now the Nancy Grace Roman Space Telescope, finding that blue photometry combined with red spectroscopy still leaves substantial degeneracies in the inferred trace-gas abundances.

\citet{CarrionGonzalez2020} explicitly accounted for
the cost of radius uncertainty: comparing retrievals with the radius fixed versus
free, they found that an unconstrained radius inflates the uncertainty
in cloud optical thickness and absorber abundance to the point where
cloudy and cloud-free atmospheres become formally indistinguishable.
Their analysis of intermediate radius priors showed that
reducing the prior uncertainty on $\Rp$ from unconstrained to roughly
$\pm10\%$ recovers most of the retrieval performance lost to the
degeneracy. \citet{CarrionGonzalez2021} demonstrated that
simultaneous multi-phase retrievals, combining small and large phase
angles, recover the radius to within about $35\%$ due to the angular structure of the phase law.

The external-prior approach substitutes independent information for the
missing radius. Mass-radius relations \citep{Muller2024} translate a
radial-velocity mass into a radius estimate, but for non-transiting
targets where only $M\sin i$ is available, the effective radius
uncertainty remains close to a factor of two, well short of the
$\pm10\%$ threshold identified by \citet{CarrionGonzalez2020}. A
transit-derived radius is the cleanest external constraint but is
available only for a small fraction of targets with favorably
oriented orbits.

In every approach noted above,  the observables
 remain extensive: at any specified phase, the reflected flux
scales as $\Rp^2$, so the radius explicitly becomes a part of the inference
and must either be retrieved internally within an assumed atmospheric
model class or supplied externally with substantial uncertainty. The
geometric albedo inferred in each case therefore inherits both the
radius prior and the assumptions of the forward model. We pursue a
different strategy: instead of constraining the radius and propagating
it into the albedo, we convert observables into intensive ratios in
which the radius-dependent normalization cancels identically, leaving
the planetary scattering state constrained independently of $\Rp$.

\subsection{Multi-epoch intensive observables}
\label{sec:intro:intensive}

The reflected intensity is an {extensive} property of the planet:
at a specified phase, it scales with the emitting area, $\propto \Rp^2$, and hence
carries the full radius-albedo degeneracy of Equation~\ref{eq:fobs}.
We note that a multi-epoch sequence, consisting of the Stokes vector of disk-integrated fluxes $\mathbf{I}_\lambda(\alpha_i) = \{I_\lambda(\alpha_i), Q_\lambda(\alpha_i), U_\lambda(\alpha_i)\}$ at wavelength $\lambda$ and phases $\alpha_i$, contains radius-free information before any absolute radius is inferred.
In contrast to extensive observables like disk-integrated flux, we define intensive observables as ratios in which the
radius-dependent normalization cancels identically, thereby constraining
the planetary state without reference to the radius. Four such
observables are available from spectrophotometry and polarimetry in a
narrow ultraviolet window:
\begin{enumerate}
\item the normalized 400 nm phase shape,
      $\shapeI=\{I_{400}(\alpha_i)/\max_j I_{400}(\alpha_j)\}$,
\item the spectral slope of the intensity at each epoch,
      $\slopeI \equiv (I_{360}-I_{400})/I_{400}$,
\item the degree of linear polarization, $\DOLP_{400}
      \equiv \sqrt{Q^2+U^2}/I$ at $400$~nm, and
\item the spectral slope of the DOLP at each epoch,
      $\slopeD \equiv (\DOLP_{360}-\DOLP_{400})/\DOLP_{400}$,
\end{enumerate}
Each is a ratio of disk-integrated quantities measured on the same
source or across the same source at different epochs: the factor
$\Rp^2$ that multiplies the Stokes vector cancels. These observables
therefore depend only on the intrinsic scattering state of the planet
and on observing geometry. In the spectrally featureless ultraviolet
regime developed in the following, the shared {atmosphere-surface} state is defined by   
$(\taufour,\rhosurf)$ and the geometry is the phase tuple
$(\alpha_1,\ldots,\alpha_N)$, where $\taufour$ and $\rhosurf$ are the optical depth of Rayleigh scattering in the atmosphere and the Lambertian surface albedo at 400~nm, respectively. 

The organization of the paper follows the observing problem. We begin
with a multi-epoch measurement vector and treat known phase as the
conditional limit in which the orbit supplies the phase tuple. We then
compare it to the unknown-phase case, where the same intensive data must
infer $(u,\rhosurf,\alpha_1,\ldots,\alpha_N)$, with
$u=\ln\taufour$.  We define the ultraviolet
forward model in section~\ref{sec:method}. In section~\ref{sec:uniqueness}, we examine how well the
multi-epoch intensive observables determine the scattering state, how
the residual state uncertainty propagates into albedo and radius, and
how these quantities change with SNR and polarimetry. Section~\ref{sec:lut}
summarizes the resulting recipe to obtain the effective planetary radius in the UV. 

This has both instrument and observing-strategy
implications for HWO. {The relevant radius-free information is delivered by near-ultraviolet color and polarization ratios.} High-SNR spectrophotometry in the 360-400~nm window, together
with linear polarimetry, therefore directly improves the ability to
separate geometric albedo from radius. This is especially important for
optically thinner, Earth-like Rayleigh columns, where residual
uncertainty in state parameters maps into a larger spread in disk brightness. When
phase is known, the known-phase map can be applied directly. In the absence of
phase information, {obtaining multiple distinct phases
becomes critical}, because phase diversity lets the same data
constrain the phase tuple as well as the {atmosphere-surface} state.

The lower bound of the wavelength range has operational significance: If the same
observing concept were shifted redward {by 40 nm to 400-440 nm}, the Rayleigh
column would be weaker by the $\lambda^{-4}$ scaling. For example, the
Rayleigh optical depth at 380 nm is $\sim$1.5 times greater than
at 420 nm. Because the radius consequence is
highly sensitive to predictable atmospheric and surface information in this regime, access
shortward of 400 nm is a scientific requirement for the cleanest
{Earth-like cases.}

\section{The ultraviolet window and the forward model}
\label{sec:method}

\subsection{Why the ultraviolet}
\label{sec:method:uv}

The UV-blue window near 360-400~nm is favorable for three
independent reasons. First, the Rayleigh scattering cross section
scales as $\lambda^{-4}$, so the atmospheric scattering optical depth
is roughly an order of magnitude larger than in the red, and the
atmosphere contributes strongly to the reflected signal even for modest
columns. {[For reference, the Rayleigh optical depth of Earth's
clear-sky atmosphere at 400 nm is $\taufour \approx 0.36$
\citep[$\taufour=0.3585$ for a standard sea-level column;][]{Bodhaine1999},
so an Earth twin inhabits the optically thin part of the regime
considered in the following. We use this value as the Earth-like reference
throughout this manuscript.]} The $\lambda^{-4}$ scaling is a detracting factor against, for example, a 400-440 nm
implementation: a measurement centered near 380 nm
samples a Rayleigh optical depth roughly 1.5 times larger than a
measurement centered near 420 nm. Since $\slopeI$, DOLP, and
$\slopeD$ are all generated by this atmospheric scattering column,
moving the window redward would reduce the intensive contrast that
breaks the radius-albedo degeneracy.

Second, most natural solid and liquid surfaces like silicate rock, soil, vegetation, and open ocean are relatively dark in the near-UV because electronic transitions in minerals and organic materials cause increased absorption at these wavelengths. Vegetation and organic-rich soils typically have reflectivities of only $\sim$ 0.03-0.08 across 360-400 nm, with tropical forests commonly near 0.04-0.06. Arid mineral surfaces are substantially darker in the ultraviolet than in the visible, with typical reflectivities of $\sim$ 0.15-0.35 \citep{HermanCelarier1997,Kleipool2008}. Open-ocean scenes are moderately reflective in this window, particularly near 380 nm, but remain far dimmer than clouds or snow. {These behaviors arise from generic condensed-matter radiative physics, and are hence not limited to Earth}: ultraviolet photons stimulate electronic-state transitions in minerals and organics, whereas large ice and water particles scatter quasi-geometrically, thus remaining bright and spectrally flat.

Third, the surfaces that
are bright in the ultraviolet form a short and well-behaved list:
snow, ice, and water or ice clouds. These are spectrally flat across
360-400 nm to within a percent
\citep{HermanCelarier1997, Kleipool2008}, because they scatter off particles much larger than the wavelength. They can
therefore be represented, to first order, as a spectrally flat Lambertian lower boundary. The Lambertian approximation is more valid in the UV because surface flux is largely diffuse owing to high Rayleigh optical depth.

We  restrict attention to a window devoid of strong
molecular absorption. On an Earth-like planet the ozone
Hartley-Huggins system suppresses reflectance shortward of
$\sim$330~nm. We work longward of this, in 360-400 nm, where the
assumption of weak absorption is defensible for the purpose of a
first-order treatment. The role of residual absorbers and of
photochemical hazes, which violate this assumption, is considered in
Section~\ref{sec:discussion}.

\subsection{Forward model and code validation}
\label{sec:method:rt}

We compute disk-integrated Stokes vectors with the GPU-accelerated
vector radiative-transfer code \vSmartMOM\
\citep{Sanghavi2014vSmartMOM, Jeyaram2022vSmartMOM}, with the
disk integral evaluated by Horak's sphere-only crescent quadrature
\citep{horak1950diffuse}. This use of \vSmartMOM\ builds on
earlier applications of the same vector radiative-transfer capability to
disk-integrated photopolarimetry of brown dwarfs and directly imaged
exoplanets \citep{SanghaviShporer2018,SanghaviWest2019,SanghaviWestJiang2021}.
Horak's change of variables places the
quadrature nodes directly inside the illuminated and visible crescent,
with final flux weights that already include the spherical area factor
and the observer projection. 
We use a $n=6$ Horak product rule, corresponding to 36 local radiative-transfer
evaluations per disk. For arbitrary observer inclination, the Horak
nodes are constructed in the natural star-observer phase frame at the
physical phase angle, $\cos\alpha=-\sin\thetai\cos\phipl$, and then
rotated into the tidally locked body frame before computing the local
solar zenith, viewing zenith, relative azimuth, and Stokes-frame
rotation; in the edge-on limit this reduces to $\alpha=\pi-\phipl$.
The atmosphere is a pure Rayleigh scatterer with
single-scattering albedo of unity and $\lambda^{-4}$ optical-depth scaling.
The surface is a spectrally flat Lambertian of scalar albedo
$\rhosurf$. We parameterize the atmospheric column by its Rayleigh
optical depth at 400 nm, $\taufour$.

\begin{figure*}[t]
\centering
\includegraphics[width=\textwidth]{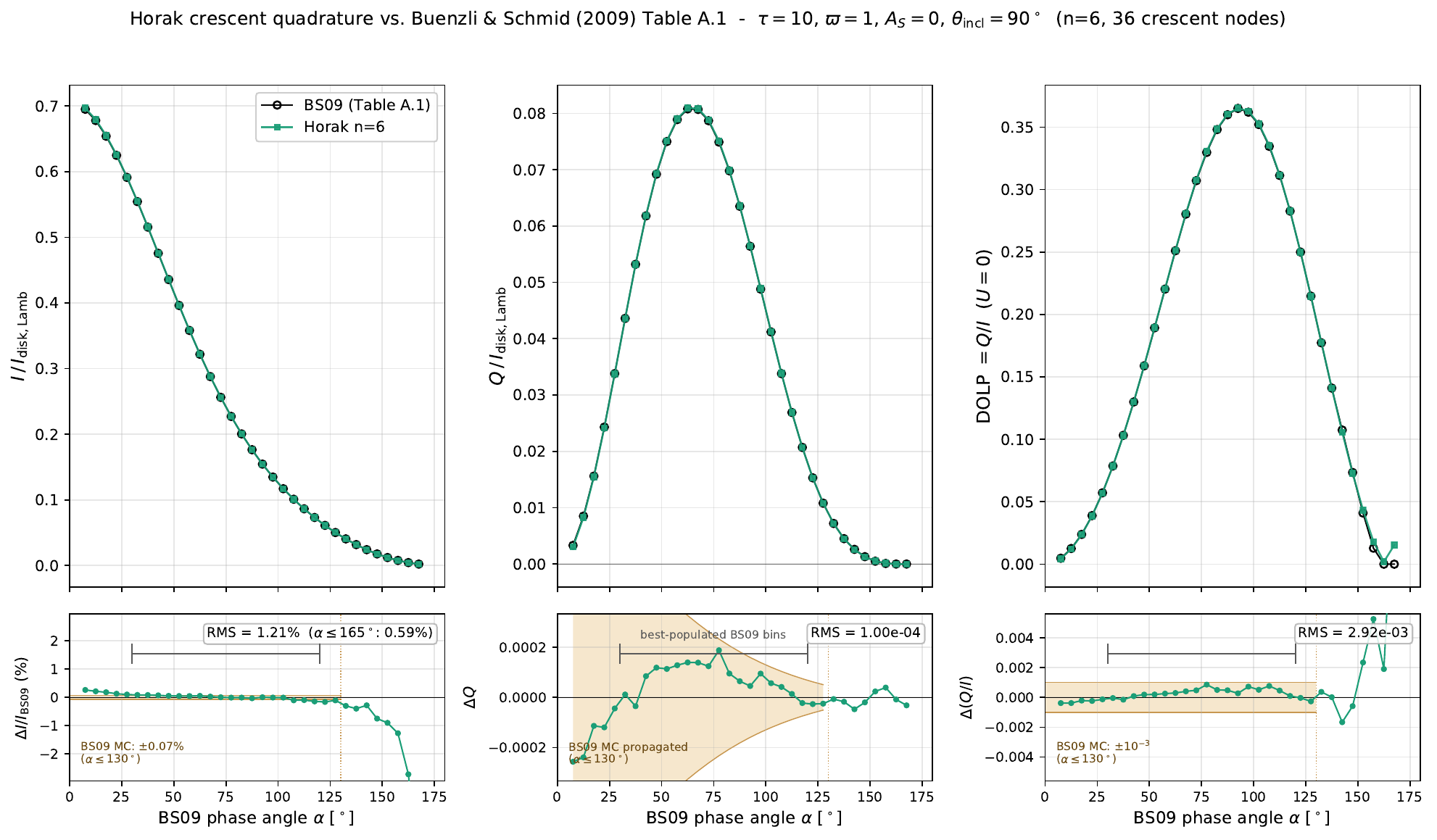}
\caption{Validation of the Horak $n=6$ crescent quadrature coupled to
\vSmartMOM\ against the \citet{Buenzli2009} (BS09) Table~A.1 benchmark
for a semi-infinite conservative Rayleigh atmosphere ($\tau=10$,
$\varpi=1$, $\rhosurf=0$, edge-on). Left to right: disk-integrated
intensity normalized to a white Lambertian disk, Stokes component
$Q$, and degree of linear polarization, each as a function of phase
angle $\alpha$. Lower panels show residuals. Amber bands mark the
formal BS09 Monte Carlo limits through $\alpha=130^\circ$ ($\pm0.07\%$
in $I$, propagated in $Q$, and $\pm10^{-3}$ in DOLP): the dotted
vertical line marks the end of that quoted limit, and the gray bracket
marks the most well-populated BS09 phase bins
($\alpha\simeq30^\circ$-$120^\circ$). The residual RMS values shown in
the panels are $1.21\%$ in $I$ over all plotted phases, or $0.59\%$
when the final crescent point is excluded, $1.00\times10^{-4}$ in $Q$,
and $2.92\times10^{-3}$ in DOLP.}
\label{fig:bs09}
\end{figure*}

Figure~\ref{fig:bs09} {compares this configuration with} the
benchmark disk-integrated polarized phase curves of \citet{Buenzli2009}
for a semi-infinite, conservatively scattering Rayleigh atmosphere over
a black surface at edge-on inclination. {We use BS09 as an
external community benchmark, so that
the comparison tests consistency between two independent
numerical approaches.} {The plane-parallel Rayleigh calculation itself is also tested against the high-accuracy benchmark tables of \citet{natraj2009rayleigh}, for which \vSmartMOM\ agrees at the tabulated precision over the tested cases. The BS09 comparison below is used specifically as a disk-integrated reflected-light benchmark.}

The three columns show the disk-integrated intensity (left), the polarized
component $Q$ (middle), and the DOLP (right) as functions of phase angle, with the
BS09 values overplotted on the \vSmartMOM\ calculation in the top panels. The lower
panels show the residuals between the two. The intensity matches the
benchmark to better than $0.6\%$ RMS over the well-conditioned range
$\alpha\le165^\circ$, the Stokes component $Q$ to $\sim$$10^{-4}$,
and the DOLP to $\sim$$3\times10^{-3}$. {The residual growth as
$\alpha\to180^\circ$ cannot be assigned uniquely to either
calculation:} near the crescent
limit the reflected signal becomes small and the contribution is
localized near the terminator and limb, so small absolute differences
are amplified in fractional residuals. We therefore quote validation
residuals for $\alpha\le165^\circ$ and choose the demonstration phases
in the following to be well clear of crescent geometry. This comparison should also be
interpreted relative to the Monte Carlo noise in the BS09 reference
calculation. \citet{Buenzli2009} report a formal model-grid limit
$\Delta(Q/I)<0.1\%$ for $\alpha=0^\circ$-$130^\circ$, corresponding to
$\Delta I/I<0.07\%$, and identify $30^\circ\lesssim\alpha\lesssim120^\circ$
as the most well-populated phase-bin range. Over that range, the
Horak/\vSmartMOM\ residuals are $0.064\%$ in intensity and
$4.4\times10^{-4}$ in DOLP, comparable to or below the BS09 noise scale.
Residuals beyond $\alpha=130^\circ$ are crescent-regime comparisons
outside the formal BS09 Monte Carlo guarantee.
A Horak-order check gives the same conclusion: increasing the crescent
rule from $n=6$ to $n=12$ (36 to 144 nodes) changes the all-phase
intensity RMS only from $1.211\%$ to $1.204\%$, leaves the
$\alpha\le165^\circ$ RMS at about $0.586\%$, and leaves the deepest
crescent residual near $-6.1\%$. {Thus, over the range relevant
to the analysis in this paper, the disk integration has effectively
converged with respect to increasing the number of Horak quadrature
points.} The endpoint discrepancy 
{ could reflect BS09 Monte Carlo sampling, phase-bin limitations, or other small
differences in implementation details.}

\subsection{The state space at fixed wavelength window}
\label{sec:method:statespace}

We computed disk-integrated Stokes vectors on a grid spanning
$\taufour \in [0, 20]${, with the Earth-like
reference value $\taufour=0.3585$ appearing near its thin end,} $\rhosurf \in \{0,0.1,\ldots,1.0\}$, and
direct-imaging phase angles sampled from $\alpha=2.5^\circ$ to
$150^\circ$, at the wavelengths
$\lambda=360$ and $400$~nm, all at edge-on inclination
$\thetai=90^\circ$ so that \vSmartMOM\'s orbital phase azimuth and the phase
angle are related as $\alpha=\pi-\phipl$. 
Intensities are normalized to a white
Lambertian disk of the assumed radius, so that the photometric ordinate
$I_{400}/\Idisk$ is the phase-dependent disk reflectivity up to the
assumed-radius normalization. At full phase, the same normalization
corresponds to the geometric albedo.

The general one-epoch state is $(u,\rhosurf,\alpha)$, with
$u=\ln\taufour$. For $N$ epochs with unknown phases, the state becomes
$(u,\rhosurf,\alpha_1,\ldots,\alpha_N)$. When the phases are known, the
phase coordinates are fixed externally, and the inferred
{atmosphere-surface} state reduces to $(\taufour,\rhosurf)$ on that
specified phase set. The symbol $\alpha$ denotes the star-planet-observer
phase angle used in the forward model; it does not require that the real
system be observed edge-on. The edge-on grid used here is a
computational simplification that lets the disk-integrated phase angle
be sampled directly for spherically shaped planets. 

\begin{figure*}[t]
\centering
\includegraphics[width=\textwidth]{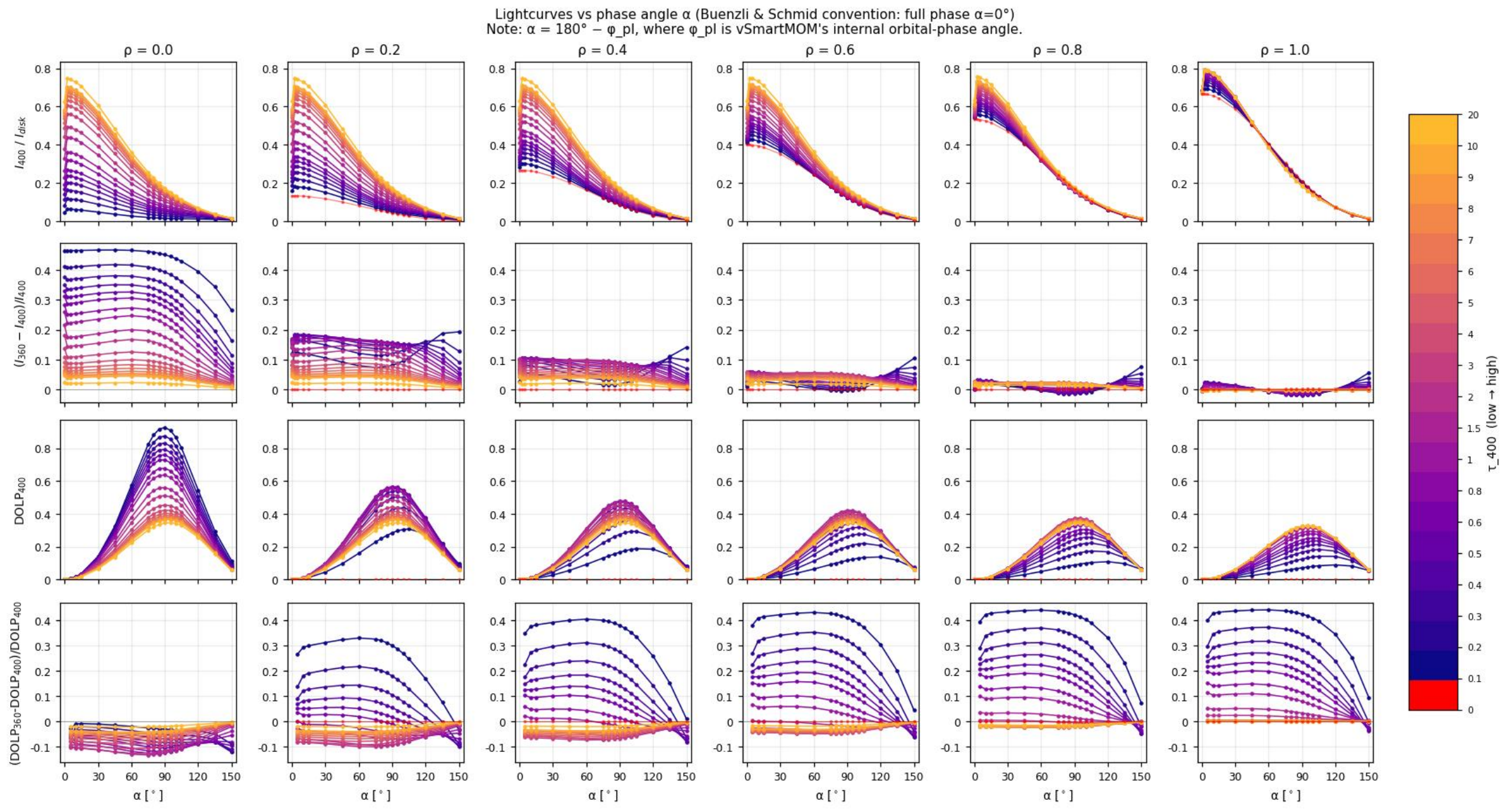}
\caption{Disk-integrated lightcurves versus phase angle
$\alpha$ over the $(\taufour, \rhosurf)$ state space, at the
edge-on inclination. Columns: surface albedo $\rhosurf$ from $0$ (dark)
to $1$ (bright). Rows: the extensive intensity $I_{400}/\Idisk$ (top) whose normalization determines the intensive observable $\shapeI$,
and the remaining three intensive observables: intensity spectral slope
$(I_{360}-I_{400})/I_{400}$, degree of linear polarization
$\DOLP_{400}$, and DOLP spectral slope
$(\DOLP_{360}-\DOLP_{400})/\DOLP_{400}$. Color encodes the Rayleigh
optical depth $\taufour$ from low (dark) to high (light). The top row
scales with the assumed disk radius, the lower three rows are
radius-free.}
\label{fig:tausweep}
\end{figure*}

Figure~\ref{fig:tausweep} shows the resulting phase curves of the four
observables for a representative subset of the surface-albedo grid. Each column corresponds to one surface albedo, from dark
($\rhosurf=0$) to bright ($\rhosurf=1$), and within each panel the
curves are colored by $\taufour$ from thin to thick Rayleigh columns,
so that each panel traces the response of an observable to the
atmospheric column for a fixed surface. The
top row is the extensive intensity $I_{400}/\Idisk$;
the lower three rows are the intensive observables $\slopeI$,
$\DOLP_{400}$, and $\slopeD$. The top row depends on the assumed disk radius (through
the normalization) and therefore carries the degeneracy, while the
lower three rows are radius-free and resolve the $(\taufour, \rhosurf)$
state. Several features are worth noting. Over a dark surface
($\rhosurf=0$, left column) the intensity climbs steadily with
$\taufour$ as the Rayleigh column brightens, while the DOLP falls
monotonically as multiple scattering depolarizes the signal. Over a
bright surface ($\rhosurf=1$, right column) the intensity is nearly
{independent} of $\taufour$, because the bright Lambertian floor
dominates, yet the DOLP rises with $\taufour$ as the polarizing
Rayleigh column builds over the unpolarized surface. The intensity is
thus least informative exactly where the polarization is most
informative.

The intensity spectral slope $\slopeI$ over a bright surface
illustrates this point: over a conservatively
scattering ($\varpi=1$) atmosphere above a white ($\rhosurf=1$) surface,
no photons are absorbed anywhere, so the total disk-integrated
reflectance is nearly wavelength-independent by construction and
$\slopeI\to0$. The only residual color comes from the angular
redistribution of photons between the atmospheric and surface paths. The
slope therefore carries Rayleigh information only when the
column is optically thin ($\taufour\lesssim1$-$2$), while  $I_{360}$ and $I_{400}$ become increasingly similar for thicker
columns. This is precisely the regime in which the
polarimetric observables, $\DOLP_{400}$ and $\slopeD$, with greater sensitivity to the scattering geometry, become the operative discriminants. 

\section{Multi-epoch conditioning, uniqueness, and radius consequence}
\label{sec:uniqueness}

Section~\ref{sec:method:statespace} establishes that a
multi-epoch sequence contains both extensive and intensive information.
The question is whether they can be used to
determine the albedo and radius uniquely enough to be useful. To study this, we introduce the
term ``consequence'' for the relative posterior uncertainty in quantities such as the inferred disk
brightness, albedo, or radius after the observables have
constrained the allowed states.

We consider two observing circumstances: known and unknown phase. In the
known-phase case, the orbit supplies each $\alpha_i$, so the inferred
state reduces to $(\taufour,\rhosurf)$. In the unknown-phase case, the
phase angles are inferred with the {atmosphere-surface} state, so the
$N$-epoch state is $(u,\rhosurf,\alpha_1,\ldots,\alpha_N)$. The
known-phase case is the conditional baseline, while the unknown-phase
case is the more general observation problem. In both cases the
normalized 400~nm lightcurve shape is an intensive observable, while
the absolute 400 nm flux scale is reserved until the final radius
inference.

The invertibility of our set of intensive parameters to a unique state
$(\taufour,\rhosurf)$ depends on the physical mapping between the state
and the observables, as well as its resilience to blurring due to
instrument noise. To study this, we use two related diagnostics: The
first is the noise-free conditioning of the observable map to help us understand the sensitivity of a given set of observables to different parts of the state space.
It is measured by
the local area metric, $\mathcal{A}$, representing the projection in observable space of a unit area of state space. Larger values of $\mathcal{A}$ indicate a well-conditioned map, while $\mathcal{A}=0$ denotes local non-uniqueness  (more details in the following). The second is the state-specific distribution of
radius consequence as a function of SNR, obtained by propagating
finite-SNR noise through the intensive observables. We also compare
these two diagnostics state-by-state, because poor local conditioning is only found to
affect the radius consequence significantly when the allowed states span different disk
brightnesses.

For one epoch, the spectropolarimetric intensive observable vector is
\begin{equation}
  \obsvec_{\mathrm{pol}}(\statevec,\alpha)
  = \bigl(\slopeI,\DOLP_{400},\slopeD\bigr)^{\top}.
  \label{eq:main_obsvec}
\end{equation}
The photometry-only subset becomes a singleton 
$\obsvec_{\mathrm{phot}}=(\slopeI)$. For a multi-epoch phase
tuple $A=(\alpha_1,\ldots,\alpha_N)$, the radius-free 400 nm shape is
\begin{equation}
  {\shapeI}_i(\statevec,A)
  =
  \frac{I_{400}(\statevec,\alpha_i)}
       {\max_k (I_{400}(\statevec,\alpha_k))} ,
  \qquad i=1,\ldots,N .
  \label{eq:main_shape}
\end{equation}
Thus the full intensive measurement vector is either
\begin{equation}
  \mathbf{y}_{\mathrm{pol}}(\statevec,A)
  =
  [\obsvec_{\mathrm{pol}}(\statevec,\alpha_1),\ldots,
   \obsvec_{\mathrm{pol}}(\statevec,\alpha_N),\shapeI(\statevec,A)]
  \label{eq:main_ypol}
\end{equation}
or, when polarimetry is unavailable,
\begin{equation}
  \mathbf{y}_{\mathrm{phot}}(\statevec,A)
  =
  [\obsvec_{\mathrm{phot}}(\statevec,\alpha_1),\ldots,
   \obsvec_{\mathrm{phot}}(\statevec,\alpha_N),\shapeI(\statevec,A)] .
  \label{eq:main_yphot}
\end{equation}
For the known-phase local maps, the inferred state is denoted
$\statevec=(u,\rhosurf)^{\top}$, with $u=\ln\taufour$. The local area
metric, $\mathcal{A}$, is computed from the Jacobian
$\jac(\statevec;\alpha)=\partial\obsvec_{\mathrm{pol}}/\partial\statevec$ (
or the corresponding photometry-only Jacobian when the DOLP rows are
omitted), as 

\begin{equation}
  \mathcal{A}
  =
  \sqrt{\det (G)},
  \label{eq:main_areametric_1}
\end{equation}
where 
\begin{equation}
G
  =
  \jac(\statevec; \alpha)^{\top}
  \jac(\statevec; \alpha).
\end{equation}
For a known set of phases, the phase-stacked metric is
\begin{equation}
  G_{\mathrm{stack}}
  =
  \sum_{i=1}^{N}
  \jac(\statevec;\alpha_i)^{\top}
  \jac(\statevec;\alpha_i),
  \qquad
  \mathcal{A}_{N\alpha}
  =
  \sqrt{\det (G_{\mathrm{stack}})}.
  \label{eq:main_areametric}
\end{equation}
This quantity is the area in observable space spanned by a unit square
in $(u,\rhosurf)$. Thus $\mathcal{A}_{N\alpha}=0$ marks local
non-uniqueness, while larger $\mathcal{A}_{N\alpha}$ indicates a better
conditioned local map.

At finite SNR, the intensive observables define weights over trial
states. Each allowed trial state has a corresponding disk reflectivity
$\Lambda=I_{400}/\Idisk$. Since the measured flux satisfies
$F_{\mathrm{obs}}\propto \Lambda\Rp^2$, the posterior spread in
$\Lambda$ gives the radius consequence,
\begin{equation}
  \frac{\sigma(\Rp)}{\Rp}
  \simeq
  \frac{1}{2}\frac{\sigma_\Lambda}{\Lambda}.
  \label{eq:main_radiusconsequence}
\end{equation}
The derivation of the area metric is given in
Appendix~\ref{app:conditioning}. The observable definitions, noise
propagation, likelihood weights, and radius-consequence calculations are
given in  Sections~\ref{app:single_phase} and \ref{app:multi_phase} of Appendix~\ref{app:phase_unknown}.

\subsection{Known phases: uniqueness, radius consequence, and polarimetry}
\label{sec:uniqueness:known}

\begin{figure*}[t]
\centering
\includegraphics[width=\textwidth]{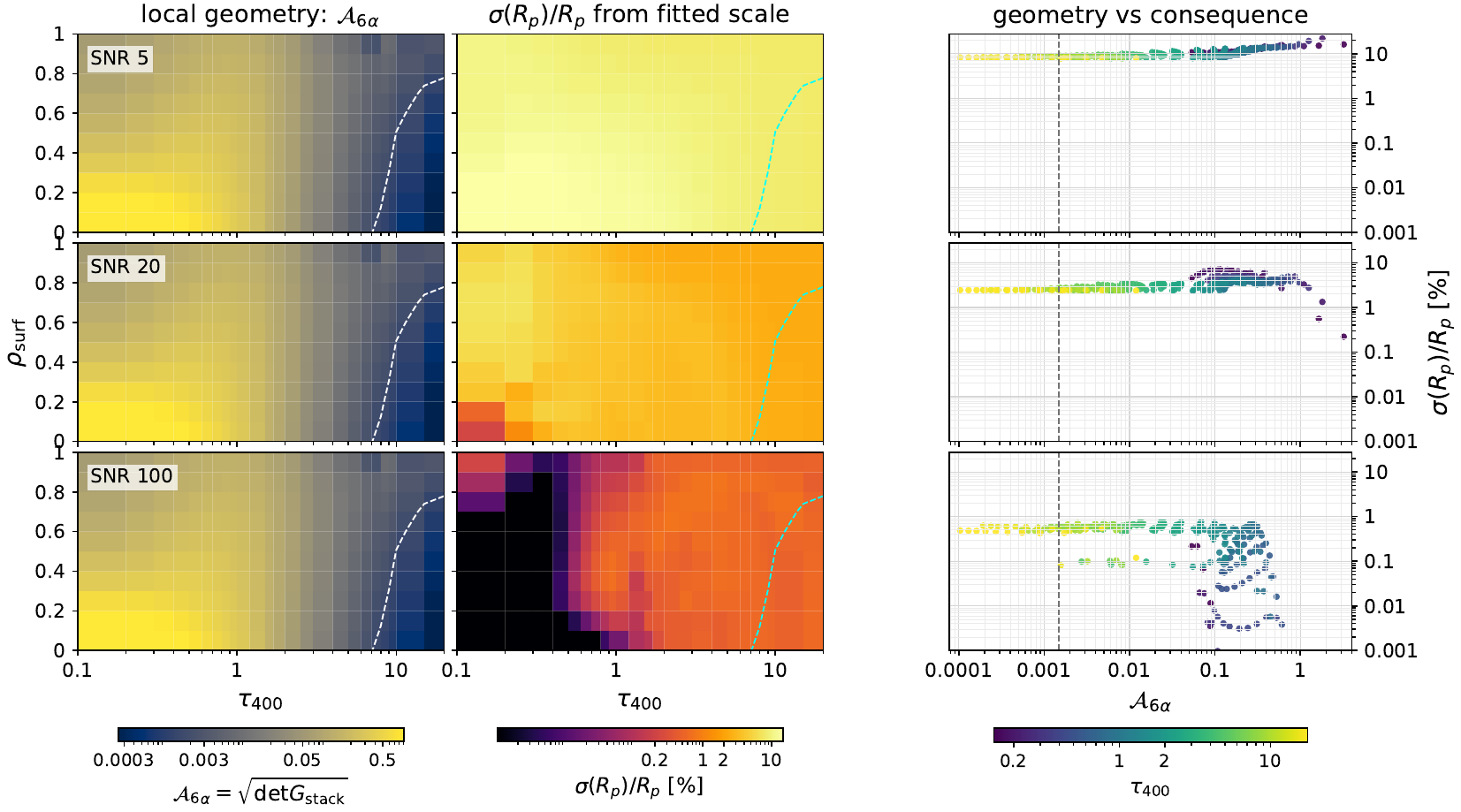}
\caption{Known-phase local state uniqueness versus radius consequence
with polarimetry. Rows show SNR $=5$, $20$, and $100$. Left: the
six-phase local area metric
$\mathcal{A}_{6\alpha}=\sqrt{\det G_{\mathrm{stack}}}$ for
$\alpha=(150^\circ,120^\circ,90^\circ,60^\circ,30^\circ,5^\circ)$.
Middle: the finite-SNR radius consequence
from the posterior spread in disk brightness,
$\sigma(\Rp)/\Rp=(1/2)\sigma_\Lambda/\Lambda$, plotted in percent. Right:
state-by-state comparison of $\mathcal{A}_{6\alpha}$ and
$\sigma(\Rp)/\Rp$, colored by $\taufour$. Dashed contours and vertical lines mark the bottom decile
of $\mathcal{A}_{6\alpha}$.}
\label{fig:unique_snr}
\end{figure*}

Figure~\ref{fig:unique_snr} shows the known-phase spectropolarimetric
result for the six-phase set
$\alpha=(150^\circ,120^\circ,90^\circ,60^\circ,30^\circ,5^\circ)$.
The left column gives $\mathcal{A}_{6\alpha}$,
repeated in each row because it is independent of SNR (see Eq.\,\ref{eq:main_areametric}).
The darkest region, representing the smallest $\mathcal{A}_{6\alpha}$ and thus highest local non-uniqueness, is the high-$\taufour$ edge, where the spectral slope
and the DOLP observables approach saturation (as seen in Fig.\,\ref{fig:tausweep}). The middle column shows
the resulting radius consequence at SNR~$=5$, $20$, and $100$. Increasing SNRs progressively deminish the uncertainties in {the inferred} radius.
While noise uniformly dominates the state space at SNR~$=5$,
a moderate SNR of 20 shows good improvements in {state resolution}, especially for thinner atmospheres over darker surfaces (Earth-like), and at SNR~$=100$,
the inversion approaches the high-precision limit. 

 At $\taufour\gtrsim1$-$2$, poor local
state-resolution often has limited radius consequence because the
allowed degeneracy direction is nearly brightness-preserving. At
$\taufour\lesssim1$-$2$, the same state uncertainty can span a larger
range of disk brightnesses, producing a more scene-dependent spread in
radius consequence. This optically thinner regime is the one most
relevant to an Earth-like Rayleigh column in the near ultraviolet, so
useful radius recovery for Earth-like cases depends especially strongly
on high-SNR measurements.

The right column plots the radius consequence of each grid cell
against its area metric, colored by $\taufour$. It is evident that the area metric is
ordered almost entirely by the optical depth: thick columns show
 small $\mathcal{A}_{6\alpha}$, while thin columns occupy
the large-$\mathcal{A}_{6\alpha}$ end. The relation between the two quantities changes
with SNR. At SNR~$=5$, the radius consequence is largest at the
large-$\mathcal{A}_{6\alpha}$ end, because the thin columns couple
their state uncertainty to a wide range of disk brightnesses, with
noise dominating every cell. As the SNR increases, this ordering
is progressively inverted: the large-$\mathcal{A}_{6\alpha}$ cells improve fastest, and
at SNR~$=100$ they reach the smallest radius consequences on the
plane, while the small-$\mathcal{A}_{6\alpha}$, high-$\taufour$ cells
improve more slowly, becoming limiting cases. The local
geometry thus sets the bounds on the precision with which planetary radius can be determined at high SNR, while the noise level
determines which part of the state plane dominates the error budget at a given SNR.

\begin{figure*}[t]
\centering
\includegraphics[width=\textwidth]{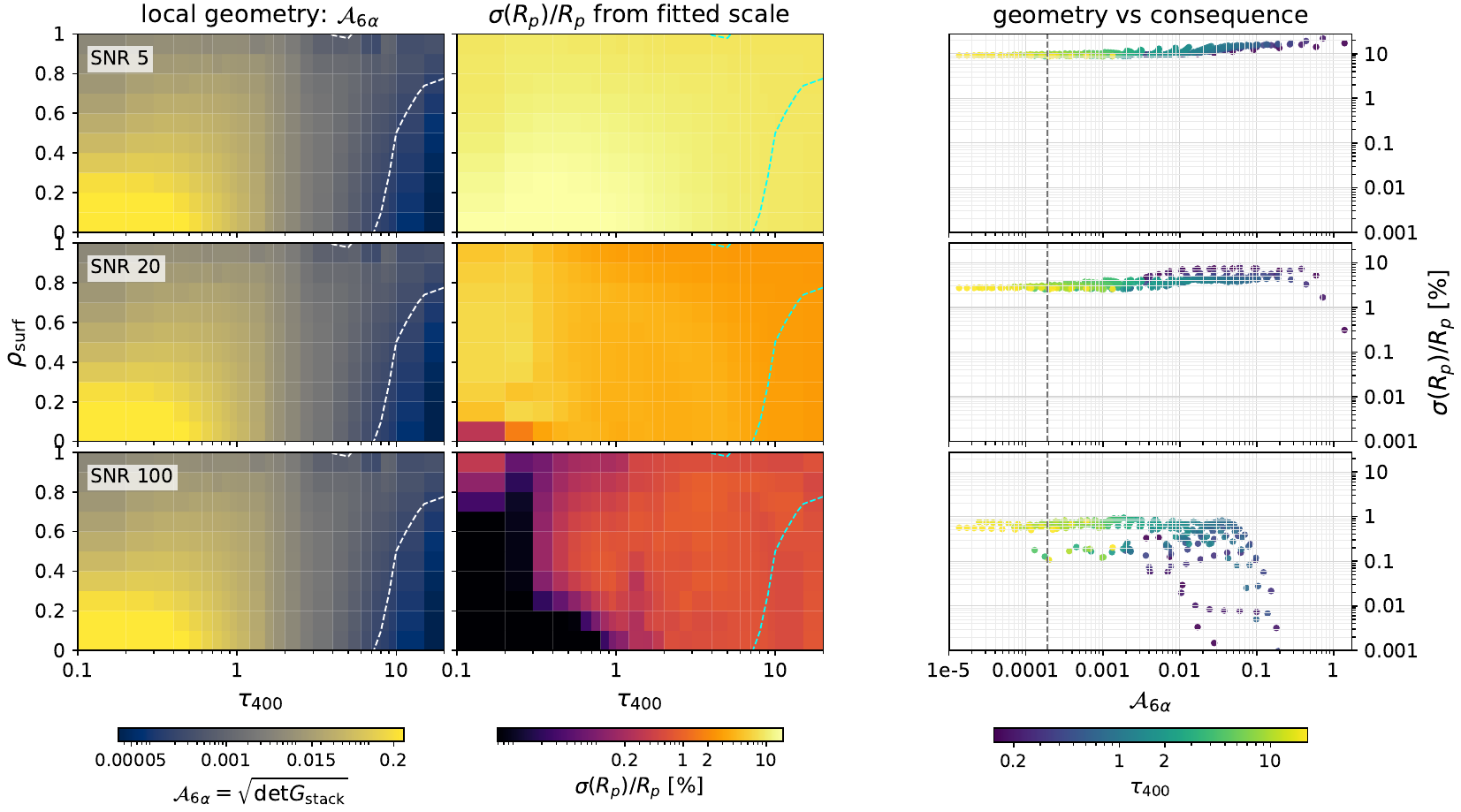}
\caption{Same as Figure~\ref{fig:unique_snr}, but without polarimetry.
The measurement vector retains the ultraviolet intensity slope and
normalized 400 nm phase shape but omits DOLP and $\slopeD$. Comparing
this figure with Figure~\ref{fig:unique_snr} isolates the information
gain supplied by the polarimetric observables.}
\label{fig:unique_snr_nopol}
\end{figure*}

Figure~\ref{fig:unique_snr_nopol} shows the corresponding values when polarimetry is not available, replacing
$\mathbf{y}_{\mathrm{pol}}$ with $\mathbf{y}_{\mathrm{phot}}$ in
Equation~\ref{eq:main_yphot}. This removes the DOLP and $\slopeD$ rows and
retains only the normalized 400 nm phase shape and the ultraviolet
color slope, so it is a strict information-loss test of the value of
polarimetry. Polarimetry increases the local area metric most clearly in the intermediate and optically thick
parts of the grid, where the intensity color has partially saturated
but the polarization still responds to the Rayleigh column and surface
reflectance. {The radius consequence, however, improves most
where the state uncertainty spans the largest range of disk
brightnesses. This occurs in the optically thin part of the grid:
at SNR~$=20$, the cell-wise polarimetric gain reaches factors of
1.3-2.2 for $\taufour\lesssim0.4$ over dark-to-moderate surfaces
($\rhosurf\approx0.1$-$0.4$), the corner that contains the Earth-like
Rayleigh column, while remaining near 1.1 over the optically thick
bulk of the plane. At SNR~$=100$ the largest gains are seen for
intermediate columns ($0.5\lesssim\taufour\lesssim2$), where the no-polarimetry
uncertainty is 2-5 times larger cell-by-cell
(see Table~\ref{tab:sec3_radius_summary}).}
For the six-phase known-phase grid, the median radius consequence with
polarimetry is $9.50\%$, $3.26\%$, and $0.45\%$ at SNR~$=5$, $20$,
and $100$, respectively. The corresponding no-polarimetry medians are
$10.13\%$, $3.76\%$, and $0.57\%$. The 90th percentiles are
$13.55\%$, $4.65\%$, and $0.64\%$ with polarimetry, compared with
$14.61\%$, $5.18\%$, and $0.78\%$ without polarimetry.
{These full-plane medians understate the polarimetric
contribution because they are dominated by the optically thick portion
of the grid, where the shape and color observables already constrain
the state, belying the regional structure of the gain summarized in
Table~\ref{tab:sec3_radius_summary}.}

\subsection{Unknown phases}
\label{sec:uniqueness:unknown}

When phase is not supplied by the orbit, the trial set includes both
the {atmosphere-surface} state and the phase tuple. The measured flux now
depends on the unknown phase as
\begin{equation}
  F_{\mathrm{obs}}(\alpha) \propto
  \Rp^2\Lambda(u,\rhosurf,\alpha),
  \label{eq:phase_flux}
\end{equation}
making it a conditional bound. The unknown-phase calculation is the conservative limit in which trial
phases may occupy the sampled phase grid. If astrometry or a
preliminary orbit fit restricts the possible phase angles without fixing
them uniquely, the likelihood should include that information as a prior
on $(\alpha_1,\ldots,\alpha_N)$. Such cases should fall between the
known-phase and fully unknown-phase limits. The likelihood, phase
marginalization (summation of posterior weight over candidate phase tuples), and radius-consequence calculation for this comparison
are given in Appendix~\ref{app:multi_phase}.

The setup is as follows:
\begin{enumerate}
\item {Known phase:} Trial phases are fixed to the observed
      set, and the posterior is over $(u,\rhosurf)$.
\item {Unknown phase:} Trial phase pairs are drawn from the
      sampled phase grid, and the posterior is over
      $(u,\rhosurf,\alpha_1,\ldots,\alpha_6)$.
\end{enumerate}

\begin{figure*}[t]
\centering
\includegraphics[width=\textwidth]{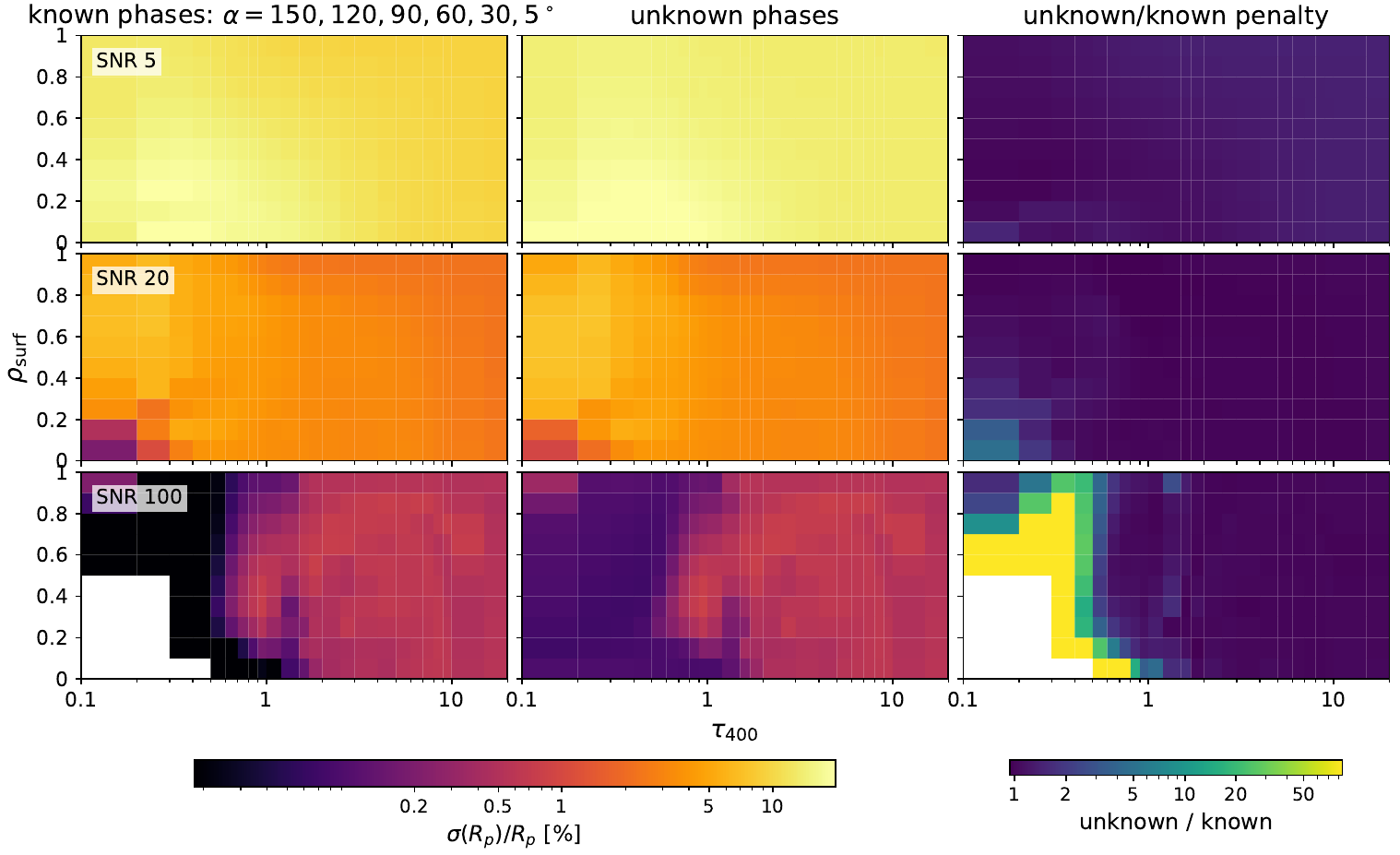}
\caption{Penalty incurred when phase is unknown, with polarimetry. The
true phase set is
$\alpha=(150^\circ,120^\circ,90^\circ,60^\circ,30^\circ,5^\circ)$.
Left: known-phase radius consequence, in which trial phases are fixed
to the true set. Middle: unknown-phase radius consequence, in which
trial phase tuples are
marginalized over the sampled grid while the normalized $I_{400}$
lightcurve shape is included with the intensive observables. Right:
unknown-phase to known-phase penalty ratio.}
\label{fig:phase_unknown_shape}
\end{figure*}

Figure~\ref{fig:phase_unknown_shape} shows the six-epoch
spectropolarimetric comparison for true phases
$\alpha=(150^\circ,120^\circ,90^\circ,60^\circ,30^\circ,5^\circ)$. The radius consequence calculations for known and unknown phases are compared side-by-side: each panel is a map over the $(\taufour, \rhosurf)$ state plane, and the three rows correspond to SNR~=5, 20, and 100. 
The left column shows the known-phase radius consequence and the middle column shows the same quantity when the phases are treated as unknown. The right column depicts their cell-by-cell ratio, representing the penalty of the
unknown-phase case against the known-phase baseline. At SNR~$=5$ (top panels),
the radius uncertainties are large throughout the state plane in both
columns, resulting in a nearly uniform penalty that never exceeds 1.5 throughout the state space.
 At SNR~$=20$ (middle row), both columns improve, most
strongly in the lower-left region of thin columns over dark surfaces
that contains Earth-like cases. The
six-epoch data now constrain the phase tuple well enough that the
penalty falls to unity over most of the plane. However, larger penalties persist in the lower-left region,
because the known-phase uncertainty falls relatively faster with increasingly SNR. At SNR~$=100$, both columns approach sub-percent levels across
 the plane. In the Earth-like region, the known-phase
uncertainty falls below the resolution of the state grid while the
unknown-phase consequence remains finite, causing the ratio to saturate.

\begin{figure*}[t]
\centering
\includegraphics[width=\textwidth]{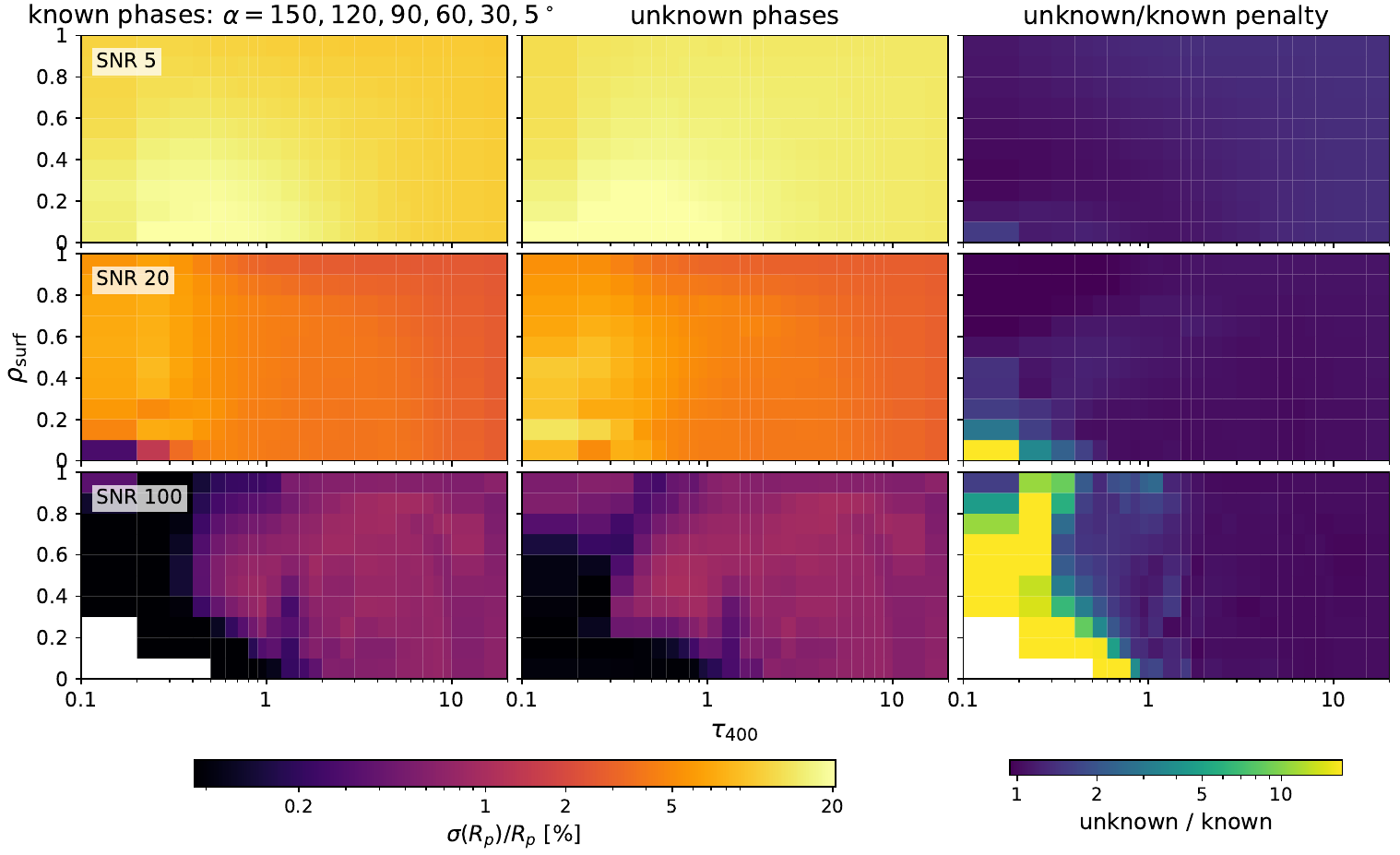}
\caption{Same as Figure~\ref{fig:phase_unknown_shape}, but without
polarimetry. The comparison retains the normalized 400 nm lightcurve
shape and ultraviolet color slope, while omitting DOLP and $\slopeD$.}
\label{fig:phase_unknown_shape_nopol}
\end{figure*}

Figure~\ref{fig:phase_unknown_shape_nopol} shows the no-polarimetry
counterpart of Figure~\ref{fig:phase_unknown_shape}. 

With six distinct phases, the unknown-phase penalty is modest
in the spectropolarimetric calculation: the median known-phase radius
consequence is $11.00\%$, $3.22\%$, and $0.45\%$ at SNR~$=5$, $20$,
and $100$, while the corresponding unknown-phase medians are
$14.00\%$, $3.20\%$, and $0.47\%$. In the majority of the state plane, the penalty is a
factor 1.27 at SNR=5, and is near unity at SNR=20 and 100.
This is because phase diversity constrains the phase tuple well enough
for the extra parameters to be included at negligible cost. The median posterior uncertainty in the {inferred} phases range from
$9.47^\circ$ at SNR~$=5$ to $1.75^\circ$ at SNR~$=20$ and
$0.22^\circ$ at SNR~$=100$. Removing polarimetry increases the
unknown-phase medians to $14.93\%$, $3.93\%$, and $0.63\%$. {The loss is again concentrated in the scientifically
valuable ranges: in the thin Earth-like corner the unknown-phase
no-polarimetry median at SNR~$=20$ is $7.22\%$ against $4.75\%$ with
polarimetry, with cell-wise gains reaching factors of 3-9. For
intermediate columns at SNR~$=100$, the median doubles from $0.29\%$
to $0.63\%$ (Table~\ref{tab:sec3_radius_summary}).} At high
SNR, some penalty maps are dominated by regions where the known-phase
denominator becomes vanishingly small - the penalty ratios should therefore not be interpreted independently of radius-consequence.

\begin{deluxetable*}{llcccccccc}
\tabletypesize{\scriptsize}
\tablecaption{{Median radius consequences $\sigma(\Rp)/\Rp$ (\%)
for the six-phase sequence
$\alpha=(150^\circ,120^\circ,90^\circ,60^\circ,30^\circ,5^\circ)$, with
and without polarimetry.}
\label{tab:sec3_radius_summary}}
\tablehead{
\colhead{} & \colhead{} &
\multicolumn{4}{c}{{With polarimetry}} &
\multicolumn{4}{c}{{No polarimetry}} \\
\cline{3-6}\cline{7-10}
\colhead{{Case}} &
\colhead{{State range}} &
\colhead{{SNR 5}} &
\colhead{SNR 10} &
\colhead{{SNR 20}} &
\colhead{{SNR 100}} &
\colhead{{SNR 5}} &
\colhead{SNR 10} &
\colhead{{SNR 20}} &
\colhead{{SNR 100}}
}
\startdata
{\multirow{4}{*}{Known phases (Sec.~\ref{sec:uniqueness:known})}} &
{full plane} &
{9.50} & {5.86} & {3.26} & {0.45} &
{10.13} & {6.33} & {3.76} & {0.57} \\
& {thin, Earth-like ($\taufour\le0.5$, $\rhosurf\le0.4$)} &
{14.47} & {9.90} & {4.00} & {$<0.01$\tablenotemark{a}} &
{15.20} & {10.28} & {5.25} & {$<0.01$\tablenotemark{a}} \\
& {intermediate ($0.5<\taufour<2$)} &
{11.03} & {6.80} & {3.64} & {0.25} &
{11.77} & {7.90} & {4.00} & {0.49} \\
& {thick ($\taufour\ge2$)} &
{8.66} & {4.60} & {2.58} & {0.56} &
{9.54} & {5.22} & {2.96} & {0.68} \\
\hline
{\multirow{4}{*}{Unknown phases (Sec.~\ref{sec:uniqueness:unknown})}} &
{full plane} &
{14.00} & {6.25} & {3.20} & {0.47} &
{14.93} & {8.14} & {3.93} & {0.63} \\
& {thin, Earth-like ($\taufour\le0.5$, $\rhosurf\le0.4$)} &
{19.20} & {11.83} & {4.75} & {0.08} &
{19.12} & {14.22} & {7.22} & {0.09} \\
& {intermediate ($0.5<\taufour<2$)} &
{14.85} & {7.12} & {3.55} & {0.29} &
{16.19} & {9.80} & {4.33} & {0.63} \\
& {thick ($\taufour\ge2$)} &
{13.38} & {5.20} & {2.62} & {0.56} &
{14.71} & {7.48} & {3.19} & {0.71}
\enddata
\tablenotetext{a}{{Below the resolution of the state grid in
both observable modes: the posterior collapses to the truth cell, so
the residual spread is not a measurable radius consequence. In the
unknown-phase case the {phase uncertainty} keeps the posterior finite, so
the corresponding entries remain quotable.}}
\end{deluxetable*}

\subsection{How many distinct phases are needed}
\label{sec:uniqueness:phasecount}

The unknown-phase penalty decreases because additional distinct phases
add angular information. {Co-adding at the same
phase would reduce noise without providing information on the angular structure
of the problem.} Distinct phases improve our sampling of the Rayleigh phase function, the
surface contribution, and the lightcurve shape, thereby reducing
the range of phase-angle tuples consistent with
the data. Improvements in the conditioning of the state result in improved
radius consequences.

\begin{figure*}[t]
\centering
\includegraphics[width=\textwidth]{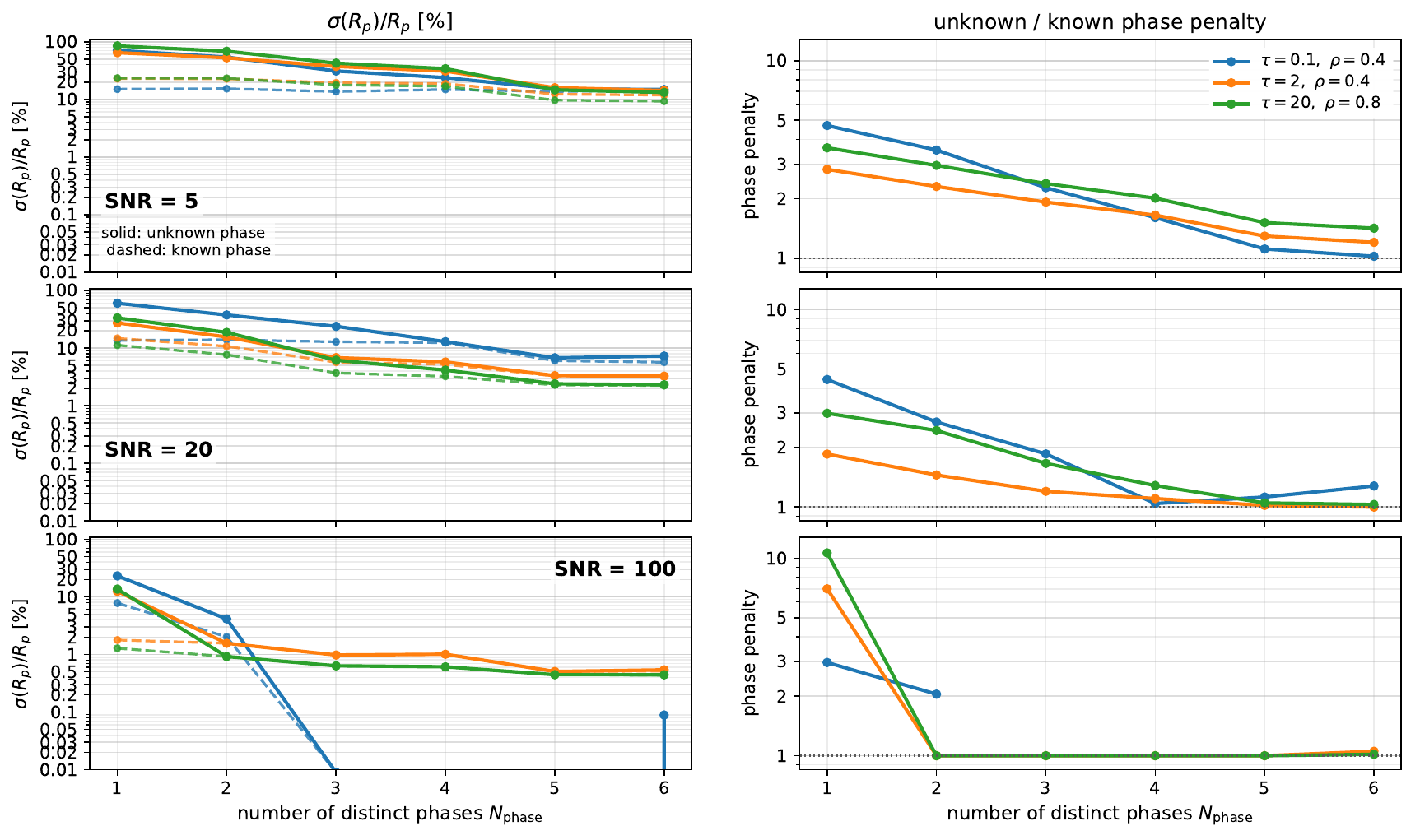}
\caption{Effect of the number of distinct phases with polarimetry, for
SNR $=5$, $20$, and $100$. The three colored curves are the
low-$\taufour$ bright-surface transition, moderate interior, and
high-$\taufour$ saturated scenes. Solid/dashed curves represent unknown/known
 phase sets. Left: absolute radius consequence. Right:
unknown-to-known phase penalty ratio. The phase design is nested,
so each step adds an epoch to the previous set.}
\label{fig:phase_count_all}
\end{figure*}

\begin{figure*}[t]
\centering
\includegraphics[width=\textwidth]{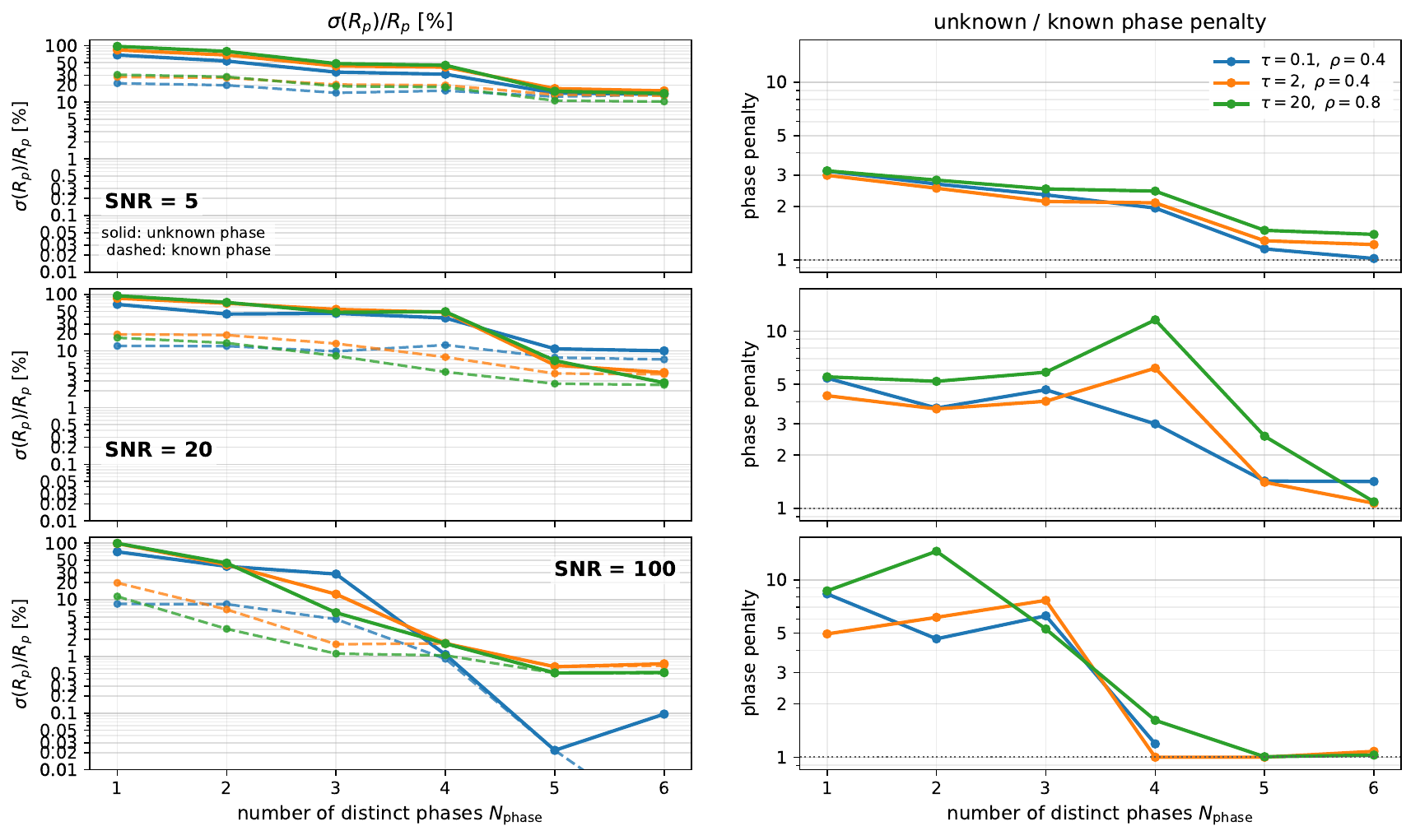}
\caption{Same as Figure~\ref{fig:phase_count_all}, but without
polarimetry. The no-polarimetry curves retain ultraviolet color and the
normalized 400 nm phase shape.}
\label{fig:phase_count_all_nopol}
\end{figure*}

To show this effect, we use nested phase sets for three scenes chosen to represent different challenging regimes: a low-$\taufour$ bright-surface transition
($\taufour=0.1$, $\rhosurf=0.4$) that represents the transition of the thin column case from dark to bright surfaces, a moderate interior scene
($\taufour=2$, $\rhosurf=0.4$) representing competing sources of intermediate brightness, and a high-$\taufour$ saturated scene
($\taufour=20$, $\rhosurf=0.8$). 
In Figures~\ref{fig:phase_count_all} (\ref{fig:phase_count_all_nopol} is its non-polarimetric counterpart), the left panel gives
the absolute radius consequence for known (dashed) and unknown phase (solid), while the right panel shows their ratio. The sequence represented by the x-axis
begins at $\alpha=90^\circ$, adds $60^\circ$, then adds $120^\circ$,
and then extends toward $30^\circ$, $150^\circ$, and $5^\circ$. Thus
increasing $N_{\mathrm{phase}}$ corresponds to the addition of observing epochs in the above order. The dashed and solid curves in the left panels converge as phases are
added. The right panels demonstrate the rate of this convergence, revealing the number of distinct epochs needed before
unknown phases stop costing radius precision.

At SNR~$=5$, phase uncertainty remains expensive for the first few
epochs, but the penalty decreases steadily. At SNR~$=20$, the penalty is
already close to unity once 5-6 distinct phases are used. At
SNR~$=100$, the known-phase bound can become so tight that some penalty
ratios look large even when the absolute radius consequence is small.
The phase diversity turns unknown phase from an
external requirement into a fitted part of the multi-epoch problem.

\section{Look-up-table inference of the planetary state}
\label{sec:lut}

In the above analysis, the forward model
defined the radius-free map between the state and the intensive observables
\begin{equation}
\label{eq:fwdmap}
(\taufour,\rhosurf,\alpha)
\;\longmapsto\;
\bigl(\slopeI,\DOLP_{400},\slopeD\bigr),
\end{equation}
as well as the extensive brightness
$\Lambda(\taufour,\rhosurf,\alpha)=I_{400}/\Idisk$. Using the intensive observables to identify the allowed region of the
forward-model grid allowed us to compute the corresponding
posterior spread in the extensive brightness $\Lambda$. Combining that
brightness with the measured absolute flux allows us to use the relationship
$F_{\mathrm{obs}}\propto \Rp^2\Lambda$ to infer the effective
ultraviolet radius. {Both the known- and unknown-phase
calculations of Section~\ref{sec:uniqueness} are lookup-table
posteriors on a discrete grid: no continuous parameter sampling, realistic
noise or forward-model error have yet been accounted for.
The results should therefore be interpreted as an idealized sensitivity study
of the information content of the intensive observables, and not as a
retrieval-performance forecast.}

Our present calculations use the discrete grid directly, with the
weights interpreted as quadrature weights over state-cell volume. In an
simulated or operational inversion, the same table can be treated as samples of a
continuous forward map. Interpolating the observables and $\Lambda$ 
over $u=\ln\taufour$, $\rhosurf$, and $\alpha$ would allow  the
problem to be adapted to the continuous state space applicable to retrievals. 

Note that our method determines the effective ultraviolet radius $\Rp(\lambda_{\mathrm{UV}})$, the radius of the equivalent reflecting
sphere in the chosen ultraviolet window. For an atmosphere over a condensed surface, this effective radius is the core radius plus the
altitude of the effective Rayleigh-scattering layer.

\section{HWO observing implications}
\label{sec:hwo_observing}

\subsection{The Habitable Worlds Observatory}
\label{sec:hwo}
We now ask what implications the above study has for the Habitable Worlds Observatory (HWO). {The flagship mission
recommended by the Astro2020 Decadal Survey \citep{NAS2021decadal} is being designed to push direct imaging beyond current limits. Where current high-contrast instruments reach planet-star flux ratios
of $\sim10^{-6}$, sufficient only for young, self-luminous giant
planets, HWO is designed to suppress starlight to the $10^{-10}$
level required to separate an Earth-size planet in reflected light
from a Sun-like host at habitable-zone separations. Its motivating
goal is to survey on the order of a hundred nearby stellar systems
and to detect and spectroscopically characterize approximately 25
potentially Earth-like worlds, searching their atmospheres for
indications of habitability and life
\citep{NAS2021decadal,Stark2024YieldMargin}.}

{Unlike transit missions, whose targets come with known
ephemerides, HWO will discover most of the planets it
characterizes.} {The current HWO observation strategy thus anticipates mostly unknown targets with poorly constrained orbits, requiring phase angles to be
inferred jointly with the atmosphere-surface state.} Multi-phase observations add angular diversity, constrain the normalized lightcurve shape, and reduce
the unknown-phase penalty. Thus, for planets without known orbits, {prioritizing distinct phase samples would provide more information than co-adding
longer at a single phase.} {Multiple revisits are in any case required to confirm candidates and establish orbits, so a multi-epoch cadence is built into the mission concept. Accordingly, we structure the observing plan as a six-epoch campaign and} assume an allocation of 72\,h on average per epoch at SNR=20, or $6\times72~\mathrm{h}=432$ h per object{, commensurate with the per-target budgets contemplated in HWO yield studies \citep{Stark2014,Stark2024YieldMargin}}.

{The instrument imposes photometric and geometric constraints on the observation campaign: The achievable planetary count
rate is determined by the telescope aperture, the optical throughput, the coronagraphic raw contrast, and astrophysical backgrounds such as
exozodiacal dust. Our exposure-time treatment follows the photon-limited formalism developed by \citet{Stark2014} for exoEarth
yield estimation, expressed in the photon-rate notation of \citet{Mennesson2024}, who also review the laboratory
starlight-suppression performance achieved to date and pathways to the required $10^{-10}$ contrast. Further, a coronagraph transmits planet light only outside an inner
working angle (IWA), the smallest angular separation from the star at which the design contrast and throughput are delivered. The IWA
scales with the diffraction limit of the telescope of diameter $D_{\mathrm{tel}}$ and is conventionally expressed as a small multiple of $\lambda/D_{\mathrm{tel}}$. We adopt}
\begin{equation}
  \theta_{\mathrm{IWA}} = \frac{3\lambda}{D_{\mathrm{tel}}}.
  \label{eq:hwo_iwa}
\end{equation}
{The projected angular separation of a planet on a
circular orbit of radius $a$, observed at phase angle $\alpha$ around
a star at distance $d_\star$, is}
\begin{equation}
  \theta_{\mathrm{p}}(\alpha)
  = \frac{a}{d_\star}\sin\alpha .
  \label{eq:hwo_sep}
\end{equation}
{A phase is observable only while
$\theta_{\mathrm{p}}(\alpha)\geq\theta_{\mathrm{IWA}}$, so the IWA
excludes the most crescent and most gibbous geometries. The extent of exclusion increases
 with increasing distance.}

{For the goal of resolving the albedo-radius degeneracy for
reflected-light planets, an ideal instrument capability would deliver
multi-epoch polarimetric observations in the near ultraviolet at high
SNR. Accessing the 360-400~nm window and measuring polarization there
improve our ability to infer albedo before using the extensive flux to
infer radius. This is directly relevant to near-UV
starlight-suppression capabilities for HWO. A short-wavelength cutoff
at or longer than 400 nm would move the diagnostic color baseline into
a region with substantially less Rayleigh optical depth, whereas
360-400 nm keeps the measurement close to the steep part of the
$\lambda^{-4}$ leverage. The need for high SNR is strongest in the
optically thinner $\taufour\lesssim1$-$2$ regime, where the allowed
state spread maps into a larger spread in disk brightness. This is
also the regime most relevant to Earth-like Rayleigh columns in the
near ultraviolet.}

In this section, we quantify the precision with which the planetary
radius can be determined for otherwise identical systems at different
distances under HWO's instrument constraints - distance degrades
the measurement through both the photon rate and through the
IWA-limited phase coverage. For each system, we select six phases that yield the widest evenly spaced phase interval satisfying both the IWA
constraint and the total-time budget. 

\subsection{Six-phase observations of hypothetical Solar twins}
\label{sec:hwo_example}
 
In the preceding sections, we described the degeneracy-breaking information
content in abstract SNR units. Now we {implement our
intensive-observable strategy for HWO observations
outlined above}. We assume
 an Earth-radius Rayleigh/Lambertian planet as our fiducial planet, with
$\rhosurf=0.2$, and $a=1~\mathrm{AU}$, observed around
synthetic solar twins at a distance, $d_\star$, of 6 and 12 pc, respectively. The instrument model assumes a telescope of diameter $D_{\mathrm{tel}}=8\,$m, two 10 nm ultraviolet bands centered at 360 and
400 nm, throughput $T=0.1$, PSF fraction
$\Upsilon=0.69$, raw contrast $10^{-10}$, three exozodis, and zero
detector read noise, dark current, and systematic contrast floor. The
exposure-time calculation follows the photon-limited {formalism
of \citet{Stark2014}, cross-checked against the photon-rate
notation of \citet{Mennesson2024}, as introduced in
Section~\ref{sec:hwo}}. The detailed count-rate equations are
given in Appendix~\ref{app:hwo_exptime}.
 
\begin{figure*}[t]
\centering
\includegraphics[width=\textwidth]{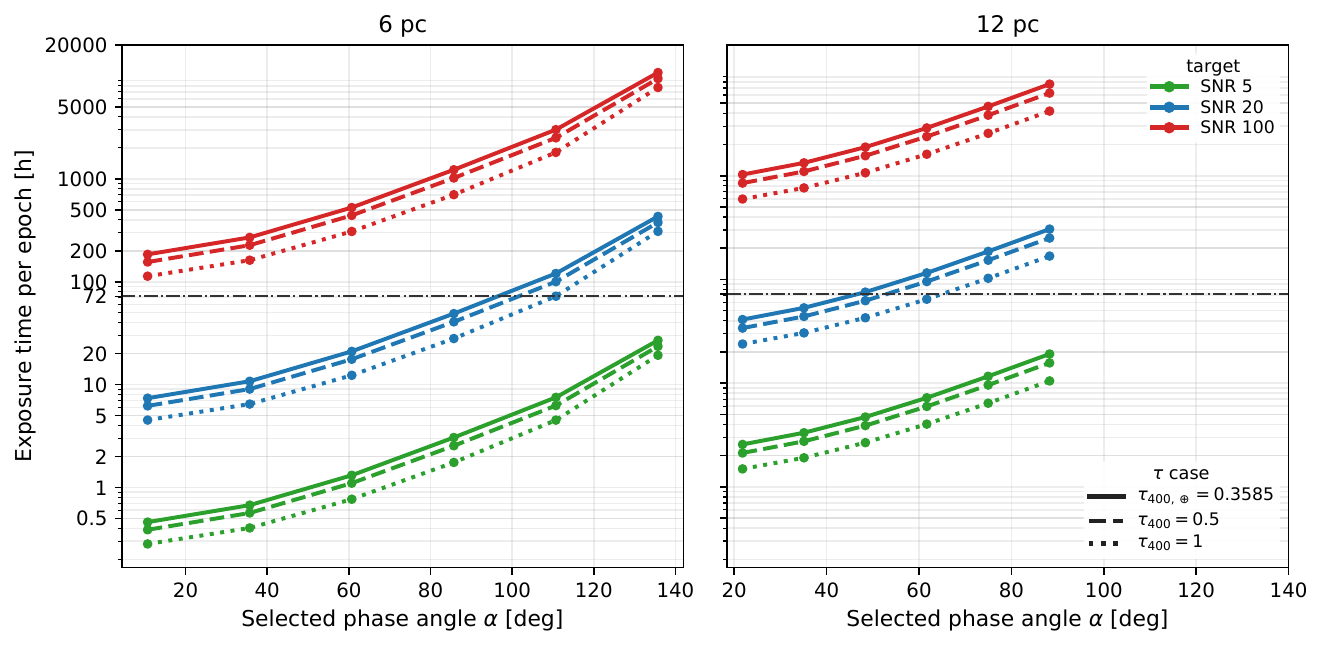}
\caption{Per-epoch HWO ultraviolet exposure times for synthetic solar
twins at 6 pc and 12 pc. Curves show the selected phase angles for
target SNR $=5$, $20$, and $100$, with line style distinguishing the
Rayleigh optical-depth cases{: solid curves show the
Earth-like column, $\tau_{400,\oplus}=0.3585$, marked by the
$\oplus$ symbol in the legend, with the optically thicker comparison
cases $\taufour=0.5$ and 1.0 dashed and dotted}. The horizontal reference marks 72 h per
epoch, corresponding to a 432 h six-epoch object budget.}
\label{fig:hwo_epoch_exposures}
\end{figure*}

{Applying the IWA constraint of Equations~\ref{eq:hwo_iwa}-\ref{eq:hwo_sep} at 400 nm, where the IWA
is most restrictive within our spectral window,} {the
resulting} $\theta_{\mathrm{IWA}}=30.94$ mas translates to low-phase limits of
$10.70^\circ$ at 6 pc and $21.79^\circ$ at 12 pc. Figure~\ref{fig:hwo_epoch_exposures} {shows the per-epoch exposure times}
for three Rayleigh columns, including the Earth-like value
$\taufour=0.3585$ at 400 nm  and the {optically thicker comparison} cases
$\taufour=0.5$ and 1.0. 
 
\begin{deluxetable*}{lccclc}
\tabletypesize{\scriptsize}
\tablecaption{Selected HWO six-phase campaigns at SNR=20.
\label{tab:hwo_phase_sets}}
\tablehead{
\colhead{Case} &
\colhead{$\alpha_{\mathrm{IWA}}$} &
\colhead{$\alpha_{\mathrm{max}}$} &
\colhead{Total time} &
\colhead{Phase set} &
\colhead{Comparison}
}
\startdata
6 pc solar twin &
$10.70^\circ$ &
$135.72^\circ$ &
432.0 h &
$135.72,110.72,85.71,60.71,35.70,10.70^\circ$ &
nearby benchmark \\
12 pc solar twin &
$21.79^\circ$ &
$88.26^\circ$ &
432.0 h &
$88.26,74.97,61.68,48.38,35.09,21.79^\circ$ &
HD 20807-like
\enddata
\end{deluxetable*}
 
Table~\ref{tab:hwo_phase_sets} summarizes the selected phase sets.
{The two distances are representative of the anticipated
HWO sample: of the 164 stars in the NASA Exoplanet Exploration Program
(ExEP) Mission Star List for HWO \citep{MamajekStapelfeldt2024}, 66
lie within 12 pc and 20 within 6 pc, so the constructed solar twins
at 6 and 12 pc cover the nearer half of the target list. The 12 pc
case is, for example, directly comparable to the solar-type target
$\zeta^2$ Reticuli (HD 20807) at 12.04 pc, while the 6 pc case is an
optimistic nearby benchmark.} At 6 pc, the target permits a much wider
phase span, but the crescent endpoint consumes most of the total time.
At 12 pc, the total-time budget compresses the campaign into a narrower,
more quadrature-centered phase range.
 
\begin{figure*}[t]
\centering
\includegraphics[width=\textwidth]{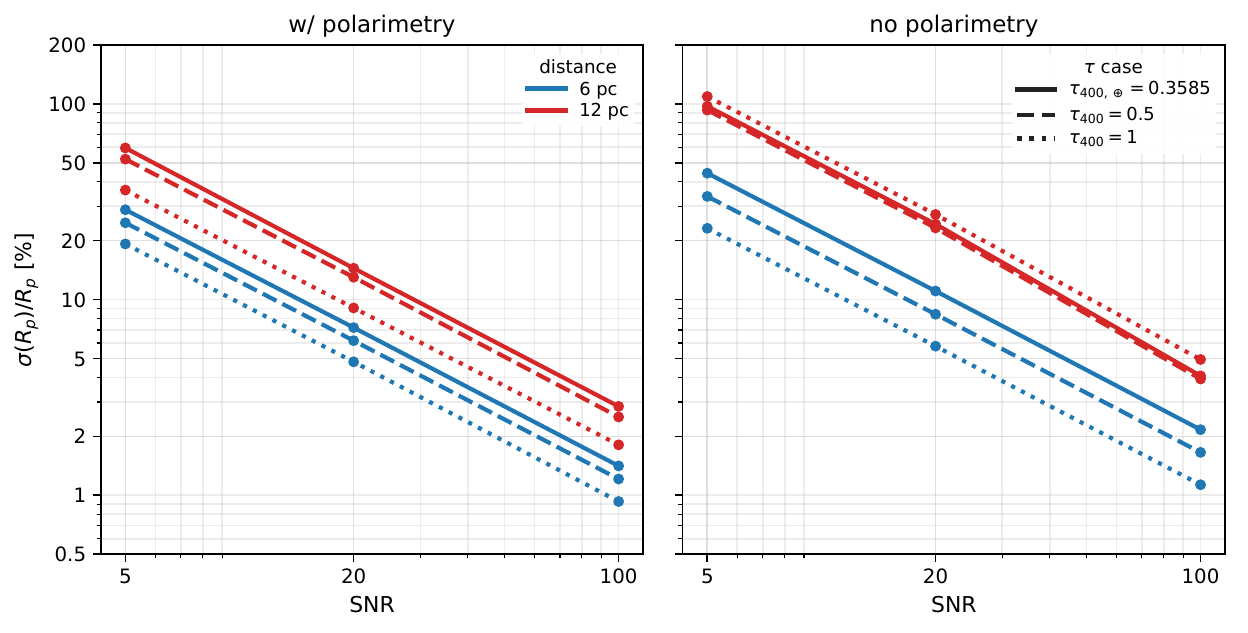}
\caption{Local-covariance radius consequences for the HWO six-phase
campaigns. The vertical axis is the fractional radius uncertainty
induced by the remaining albedo/phase degeneracy after fitting the
radius-free observables. Left: full spectropolarimetry, using
$\slopeI$, $\DOLP_{400}$, $\slopeD$, and the normalized 400 nm
lightcurve shape. Right: no-polarimetry limit, using only $\slopeI$ and
the normalized 400 nm shape. Blue and red curves show the 6 pc and
12 pc solar-twin examples. Line style distinguishes the considered
Rayleigh optical depths, with the solid curves showing the
Earth-like column $\tau_{400,\oplus}=0.3585$ ($\oplus$ in the
legend). The phase angles are fixed to those selected
from the $\taufour=1$, $\rhosurf=0.2$, SNR=20 total-budget calculation.
Lower Rayleigh optical depths are shown {at the same phases rather than
being re-optimized}, so their total SNR=20 campaign times can exceed the
432 h design budget.}
\label{fig:hwo_radius_consequences}
\end{figure*}

In Figure~\ref{fig:hwo_radius_consequences}, we show the radius
consequence, $\sigma(\Rp)/\Rp$, of the six-epoch HWO observations described above as a
function of SNR. These calculations differ from the diagnostic
degeneracy maps shown earlier in the paper: here, each point is obtained
{from an explicit inversion of synthetic observations, rather than by
comparing discrete grid cells.} For each distance, Rayleigh column, and
SNR, we generate {noise-free realizations of the}
synthetic observables (the model observables are left unperturbed) at the retained six
phases, including both the per-epoch intensive observables and the
normalized 400 nm lightcurve shape. We evaluate the Horak/vSmartMOM
lookup table with cubic interpolation between grid points and
iteratively minimize a full-covariance weighted residual,
$\chi^2 = (\mathbf{y}_{\rm mod}-\mathbf{y}_{\rm obs})^T
\mathbf{C}^{-1}(\mathbf{y}_{\rm mod}-\mathbf{y}_{\rm obs})$, taking into account the exact form of HWO noise prescribed by \citet{Stark2014, Mennesson2024}. Employing a multistart search over the state space, the
lowest-cost solution is used to compute the local posterior
covariance. The details of this inversion are given in
Appendix~\ref{app:hwo_inversion}. The {inferred} state is
\begin{equation}
  \mathbf{x}
  =
  [\ln\taufour,\rhosurf,\alpha_1,\ldots,\alpha_6] ,
  \label{eq:hwo_state}
\end{equation}
with the phase prior restricted to the HWO-feasible interval for the
corresponding system. Choosing the best-fitting intensive state, we propagate the local covariance through the model brightness to obtain
the radius uncertainty implied by the remaining albedo-phase degeneracy. The left panel of Figure~\ref{fig:hwo_radius_consequences} represents the
polarimetric case, using the full spectropolarimetric observable set, and
the right panel represents the corresponding no-polarimetry case, in
which DOLP and the DOLP spectral slope are removed. The smaller values of the left panel indicate
that the retained observables leave less uncertainty in the disk
brightness, and hence in the radius inferred from the absolute flux.
The best fits recover the truth to numerical precision in all reported
cases. 
 
\begin{deluxetable*}{lccccccccc}
\tabletypesize{\scriptsize}
\tablecaption{Local-covariance radius consequences for retained HWO
phase sets{, showing the full spectropolarimetric observable
set alongside the intensity-only limit at each SNR}.
\label{tab:hwo_radius_consequences}}
\tablehead{
\colhead{} &
\colhead{} &
\multicolumn{4}{c}{Spectropolarimetric $\sigma(\Rp)/\Rp$ [\%]} &
\multicolumn{4}{c}{{No-polarimetry $\sigma(\Rp)/\Rp$ [\%]}}\\
\cline{3-6}\cline{7-10}
\colhead{Case} &
\colhead{$\taufour$} &
\colhead{SNR 5} &
\colhead{{SNR 10}} &
\colhead{SNR 20} &
\colhead{SNR 100} &
\colhead{{SNR 5}} &
\colhead{{SNR 10}} &
\colhead{{SNR 20}} &
\colhead{{SNR 100}}
}
\startdata
6 pc solar twin  & 0.3585 & 28.77 & {14.39} &  7.19 & 1.41 & { 44.23} & {22.11} & {11.06} & {2.17} \\
6 pc solar twin  & 0.5    & 24.68 & {12.34} &  6.17 & 1.21 & { 33.66} & {16.83} & { 8.41} & {1.66} \\
6 pc solar twin  & 1.0    & 19.24 & { 9.62} &  4.81 & 0.93 & { 23.12} & {11.56} & { 5.78} & {1.13} \\
12 pc solar twin & 0.3585 & 59.51 & {29.75} & 14.46 & 2.85 & { 97.22} & {48.61} & {24.30} & {4.08} \\
12 pc solar twin & 0.5    & 52.17 & {26.08} & 13.01 & 2.51 & { 92.93} & {46.46} & {23.23} & {3.93} \\
12 pc solar twin & 1.0    & 36.27 & {18.13} &  9.07 & 1.81 & {108.96} & {54.45} & {27.21} & {4.94}
\enddata
\end{deluxetable*}
 
Distance affects the inference through
phase coverage as well as exposure time. The 6 pc case permits
{a broad phase span, including a deeper crescent, whereas the 12 pc case
is forced to adopt a narrower, more quadrature-centered window} due to 
IWA and time-budget constraints. The resulting loss of angular coverage makes the
12 pc radius consequence larger, especially when polarimetry is removed.
For the fiducial $\taufour=1$ case at SNR=20, the six-phase
radius consequence with polarimetry is $4.81\%$ at 6 pc and $9.07\%$ at
12 pc, while without polarimetry we get $5.78\%$ and $27.21\%$.
For an Earth-like Rayleigh optical depth estimate of
$\taufour=0.3585$ at 400 nm, using the same phase angles, the
corresponding full-observable values are $7.19\%$ and $14.46\%$, with
total SNR=20 campaign times of 639 h and 777 h. Without polarimetry, the
Earth-like cases yield $11.06\%$ and $24.30\%$. {At SNR $=10$, the Earth-like radius
consequences are 14.4\% at 6 pc and 29.8\% at 12 pc with polarimetry,
and 22.1\% and 48.6\% without, while the total campaign times shrink
to $\sim$160 h and $\sim$194 h
(see Table~\ref{tab:hwo_radius_consequences}). A six-epoch SNR $=10$
campaign therefore fits within practical time allocations while still
constraining the radius well below the factor-of-two uncertainty of
external mass-radius priors for non-transiting planets
\citep{Muller2024,CarrionGonzalez2020}.} It should be noted that these values, obtained {from unperturbed model observables}, are not end-to-end HWO performance predictions. They simply show how the proposed
intensive-observable method couples to a plausible exposure-time scale
in which phase coverage, distance, polarimetry, and Rayleigh optical
depth affect the radius consequence.
 
\subsection{Limitations of the {noise-free-realization} HWO target assessment}
\label{sec:hwo_noiseless_limits}

{The HWO simulations above are noise-aware but
realization-free: the covariance matrices propagate the full
astrophysical noise budget of Appendix~\ref{app:hwo_exptime},
including planetary shot noise and the residual-starlight, zodiacal,
and exozodiacal backgrounds at the stated SNR, but the synthetic
observables themselves are evaluated at their exact model values and
are never perturbed by random noise. The results should
therefore be interpreted as a target-assessment exercise rather than
as a complete retrieval-performance forecast.} They are useful for contrasting
different observing choices like SNR, number of phases, and the inclusion of polarimetry vs. intensity-only measurements on a uniform quantitative basis. Our simulations identify configurations that preserve more radius information and
observation compromises that could be costly. The numerical radius consequences,
however, are more qualitative than quantitative. They are local
covariance consequences around {unperturbed} synthetic observations, and
thus do not include the scatter, bias, multimodality, or failure
modes that would appear in noisy realizations of the measurement. A
more exact assessment would require a suite of full synthetic retrievals
against known truth values of $\Rp$, $\taufour$, $\rhosurf$, and phase,
with realistic noise injected into the simulated observations. Such an ensemble would allow the reported
radius consequence to be evaluated as an actual retrieval error
distribution, {rather than as the local diagnostic used here.}

{A further instrument-level consideration is spectral
bandwidth. The two 10 nm bands adopted here are narrow relative to
the $\sim$20\% instantaneous bandwidth anticipated for HWO
coronagraph channels \citep{Mennesson2024}. Wider bands within
360-400 nm would raise the photon rates and shorten the exposure
times, but at the cost of partially averaging the intensity and DOLP
spectral slopes across the steep $\lambda^{-4}$ gradient that
generates them. Quantifying this sensitivity-versus-leverage trade
requires band-integrated forward modeling and is deferred to
follow-up work.}

{The path from this idealized study to a synthetic, and
eventually real, retrieval is nevertheless well-defined. The free
phase tuple of Section~\ref{sec:uniqueness:unknown} could obtain priors from
an orbit-parameterized tuple anchored to known epoch times,
or from the astrometry that the same imaging epochs
deliver. The look-up table that discretizes our forward model would be interpolated continuously
in $(\taufour,\rhosurf,\alpha)$ with explicit grid-convergence tests.
This would also allow us to test and quantify the forward model error caused by departures of real planets from homogeneous
Rayleigh-Lambertian assumptions.}

{\subsection{UV spectropolarimetry: tradeoffs}}
\label{sec:hwo_tradeoffs}

{The near-ultraviolet coverage and polarimetry that define the
observing strategy developed above
involve significant engineering costs for HWO. Extending
high-contrast coronagraphy shortward of 400~nm tightens requirements
on wavefront stability, optical surface quality, scattered light, and
polarization aberrations, because coronagraphic contrast at the
$10^{-10}$ level becomes progressively harder as $\lambda$ decreases
\citep{Balasubramanian2011,Breckinridge2015,Kim2025SCoOB}. Polarimetry
demands instrumental real estate (a Wollaston prism, polarization modulator,
or equivalent), a calibration burden (instrumental polarization known
to $\lesssim10^{-4}$ for absolute reflected-light studies), and a
photon-budget cost, since the Stokes $I$ signal is split between two
orthogonal channels.} Can these costs be justified by the corresponding scientific gains? 

{A two-decade body of work argues
that polarimetry delivers information that flux measurements cannot.
\citet{GoodisGordon2025} show that an Earth-like planet could be characterized in polarized
and unpolarized light throughout its geological history, noting that
flux-only strategies exploit only a fraction of the information content
of reflected starlight. \citet{Chubb2026} arrive at the same conclusion for
the HWO instrument suite, and \citet{Kane2026} develop the closely
related spectral, polarimetric, and UV case for Venus-like worlds.
These build on extensive reflected-light polarimetry
studies \citep[e.g.,][]{Stam2008,KaralidiStam2012,TreesStam2019,
Groot2020,TreesStam2022} and on the community white paper to the
National Academies' exoplanet science strategy
\citep{MillarBlanchaer2018}. The UV window is likewise already a
recognized HWO science requirement for rocky-planet biosignatures:
ozone and related species have their strongest features in the
near-UV, accessible at modest spectral resolution
\citep{Damiano2023,Schlecker2025}. The stellar NUV inputs needed
for such observations are being assembled for HWO target systems
\citep{Bhattacharyya2026}. The intensive-observable method presented by this
paper makes a distinct contribution to this list by using the NUV window to break the
radius-albedo degeneracy.}

{ Solar System research points at two important precedents in which
exactly this combination of disk-integrated polarimetry with a
near-UV spectral window resolved compositional and structural
ambiguities that could not be resolved by flux measurements. The sulfuric-acid
composition and narrow $\sim$1~$\mu$m size distribution of the Venus
cloud droplets were established from ground-based multi-wavelength
polarization phase curves in which the 365~nm band constrained the
refractive-index \citep{HansenHovenier1974}, four years
before in-situ confirmation. Pioneer Venus polarimetry at 270 and
365~nm subsequently revealed the planet's submicron upper haze
\citep{Kawabata1980}. For Titan, ground-based polarimetry
\citep{Veverka1973,Zellner1973} and UV photometry \citep{Caldwell1975}
established the aerosol haze. The incompatibility of Pioneer~11 and Voyager~2 photopolarimetry at 264
and 750~nm \citep{TomaskoSmith1982,West1983Titan} with any spherical
particle model forced the fractal-aggregate description
\citep{WestSmith1991} that Cassini-Huygens confirmed in situ more
than two decades later \citep{Tomasko2008}.}

{The required capabilities have been
studied and are advancing on the engineering side as well: POLLUX, developed for LUVOIR, demonstrated
a formally costed high-resolution spectropolarimeter design covering
100-400~nm \citep{Bouret2018POLLUX}. A UV integral-field spectrograph
has been added to the HWO exploratory analytic cases EAC4 and EAC5 in
response to community science case development documents
\citep{Vieira2026PyISH} and laboratory demonstration of UV
coronagraphy at HWO-relevant contrast is underway \citep{Kim2025SCoOB}.}

{Within the current HWO architecture development, the exploratory analytic
cases \citep{Liu2026EAC} baseline the main coronagraph at
400-1700~nm, but include an unpopulated instrument bay that can
accommodate a to-be-determined instrument, for example a UV
coronagraph or a camera coupled to a UV starshade. A near-UV starshade
is under consideration as a second-generation capability at the
30-35~m scale consistent with current starshade technology
\citep{Liu2026arch}. HWO is also designed for robotic servicing,
through which UV capabilities could be implemented after launch. The
present work adds a unique science case for such a UV
starlight-suppression instrument: planetary-model-independent recovery of the
planet radius and geometric albedo of habitable-zone targets, at
the exposure-time scales quantified in
Section~\ref{sec:hwo_example}.}

\section{Discussion}
\label{sec:discussion}

\subsection{Assumptions underlying our method}
\label{sec:discussion:assumptions}

Our method is based on a set of first-order assumptions listed below. Future work will test their limits by quantifying forward modeling biases on the precision of radius retrievals.
\begin{enumerate}
\item \textbf{Spectrally flat Lambertian surface.} We represent the
      surface as a Lambertian albedo $\rhosurf$, constant across
      360-400~nm. The spectral-flatness assumption is supported for the relevant bright surfaces:  snow, ice, and clouds are flat in the ultraviolet to a
      fraction of a percent \citep{HermanCelarier1997, Kleipool2008},
      while the contribution of any spectral structure in dark surfaces is
      suppressed by their small absolute reflectivity. The Lambertian
      assumption could be violated by specular
      surfaces and strongly
      non-Lambertian ices, which need to be addressed separately.
\item \textbf{Homogeneous disk.} We treat the surface and atmosphere as
      laterally uniform over the visible hemisphere. A real planet
      presents a heterogeneous disk, with continents, oceans, and variable
      cloud cover. The rotational and orbital modulation of a heterogeneous disk will be an important extension of this work.
\item \textbf{Clouds and hazes as a thick Lambertian boundary.} We
      approximate optically thick clouds as a bright, spectrally flat
      Lambertian surface placed at the cloud top. Though this captures the
      first-order brightening, it cannot account for the polarizing properties of cloud
      particles or the forward-scattering signature of sub-micron
      photochemical hazes. Hazes can often be absorbing in the
      ultraviolet, violating the weak-absorption assumption of
      Section~\ref{sec:method:uv}, while also altering the
       phase law, each requiring updates to our forward model. 
\item \textbf{Weak absorption in 360-400~nm.} We assume the window is
      free of strong molecular absorption. On an Earth-like planet this
      requires working longward of the ozone Hartley-Huggins ($\sim$200-370~nm) system
      \citep[e.g.,][]{Gorshelev2014,Huang2019}.
      This is not, however, a universal
      guarantee of a clean ultraviolet window. Other atmospheric states
      can introduce absorption in or near this interval: $\mathrm{SO_2}$
      in volcanically active atmospheres
      \citep{Vandaele1994SO2,SpragueJoens1995,Shinohara2008},
      $\mathrm{NO_2}$ from lightning or high-NOx chemistry
      \citep{Vandaele1998NO2,Logan1983,Tie2002}, BrO and related
      halogen species from marine or sea-salt photochemistry
      \citep{Wahner1988,Sander2003,SaizLopez2004}, and sulfur-bearing gases or
      photochemical hazes in reducing or volcanically supplied
      atmospheres. In addition, moving the diagnostic window longward of
      400 nm does not solve this problem for Earth-like planets, because
      the ozone Chappuis band introduces additional visible absorption
      structure \citep{Brion1998,Huang2019} simultaneously as Rayleigh scattering weakens.
      Identifying clean intervals, or explicitly retrieving trace
      absorbers with the Rayleigh/surface state, will thus be
      planet-specific.
\end{enumerate}

\subsection{Existing degeneracy-breaking strategies}
\label{sec:discussion:relation}

{The method proposed here is complementary to the retrieval-based approaches of
Section~\ref{sec:intro:sota}. Hitherto approaches constrain the radius
(internally or via external priors) and propagate it into the albedo, while
ours constrains the albedo directly, through radius-free observables,
and propagates it into the radius.} The two are independent and could be
combined: the intensive-observable inference provides a
simple prior on $(\taufour, \rhosurf)$ that a full retrieval can
further refine. The multi-phase strategy of \citet{CarrionGonzalez2021}
 already exploits phase-angle diversity, but while its flux-based retrieval includes radius as a fit parameter, the intensive observables could remove the radius from the inference entirely.


{\subsection{Toward the full state space}}
\label{sec:discussion:future}
{This conceptual study can be extended in the future to include: (i) a formal probabilistic inversion that propagates realistic photometric and polarimetric noise into a posterior over $(\taufour, \rhosurf)$ at known phase, or over $(\taufour,\rhosurf,\alpha_1,\ldots,\alpha_N)$ when the phases are unknown, (ii) relaxation of the homogeneous-disk assumption, using rotational light curves to constrain surface and cloud heterogeneity, (iii) explicit treatment of photochemical hazes and non-Lambertian surfaces, and (iv) the connection of the recovered effective ultraviolet radius to the core radius, via the chromatic dependence of the effective scattering altitude and the separate determination of atmospheric scale height from pressure-broadened bands where available. Each of these cases presents an avenue for the application of the clean intensive/extensive separation of observables established in this work to more realistic/complex scenes.}

\section{Conclusions}
\label{sec:conclusions}

We have shown that the radius-albedo degeneracy in directly imaged
exoplanets can be broken using multi-epoch observables that are
independent of the planetary radius. The absolute disk-integrated
reflected intensity is an extensive observable that carries the radius-albedo degeneracy. The radius dependence cancels through the construction of intrinsic variables like
the normalized 400 nm phase shape, the intensity spectral slope, the degree
of linear polarization, and the DOLP spectral slope. In a spectrally featureless
ultraviolet window where Rayleigh scattering dominates and natural
surfaces are either dark or spectrally flat, these intensive observables
depend only on the Rayleigh optical depth, the surface albedo, and the
phase tuple.

Using \vSmartMOM, validated for 1D radiative transfer against \citet{natraj2009rayleigh} and disk-integrated radiative transfer against the \citet{Buenzli2009} benchmark,
we quantified the albedo and radius consequence of this map under two
observing circumstances: When the phases are known, the six-phase
spectropolarimetric map is locally unique over most of the state plane,
and the radius consequence decreases strongly from SNR~$=5$ to
SNR~$=100$. Treating phase as unknown is more conservative, but the
six-phase normalized lightcurve shape substantially reduces the
unknown-phase penalty. No-polarimetry simulations show the
information loss incurred when DOLP and $\slopeD$ are unavailable. 

For HWO's Earth-like reflected-light characterization, high-SNR spectrophotometry and polarimetry in the 360-400~nm window provide more precise and scientifically more useable radius inferences than photometry alone.
Particularly for optically thinner, Earth-like Rayleigh columns, 
residual state uncertainty produces a larger spread in disk brightness. The strongly diminishing strength of Rayleigh scattering with increasing wavelength presents a strong case for maintaining a window shortward of 400~nm.
For targets without orbits precise enough to supply the phases, multi-phase sampling provides the phase diversity needed to determine the phase law
simultaneously with the atmosphere-surface state.

A robust validation path for our method may be supplied by transiting rocky
planets whose radii are known independently. Recent
atmospheric-retention tests span both positive and negative cases:
helium escape from the habitable-zone rocky planet LHS~1140b (15 pc) suggests
that it has retained an atmosphere \citep{Dittmann2017,Cherubim2026},
{whereas thermal-emission measurements of LHS~3844b (14.9 pc) and TRAPPIST-1b/c (12.4 pc)
have been used to rule out thick atmospheres in
highly irradiated rocky systems}
\citep{Kreidberg2019,Greene2023,Zieba2023}. {Such objects are not direct
HWO analogues.} LHS~1140b, for example, is a transiting
M-dwarf planet at about 15 pc and has a small projected separation.
They could, however, represent valuable data-points in terms of
atmospheric-evolution, because their independently
known radii allow the radii {inferred} from HWO's reflected-light
observables to be compared against an external truth. Extending the HWO
calculation toward Sun-like systems at distances approaching 15 pc
would thus connect the radius-{determination} method to the broader
question of which rocky planets retain Earth-like atmospheres over
geological time.

Our first-order treatment using assumptions of
spectrally flat Lambertian surfaces, a homogeneous disk, and clouds
approximated as a thick Lambertian boundary define the path toward a
fuller exploration of the exoplanet state space.

\appendix

\section{Conditioning and radius-consequence calculations}
\label{app:phase_unknown}

This appendix gives the mathematical steps used for the conditioning,
known-phase, and unknown-phase comparisons in
Section~\ref{sec:uniqueness}. The
reader-facing phase angle is
\begin{equation}
  \alpha = 180^\circ-\phipl ,
  \label{eq:app_phase}
\end{equation}
where $\phipl$ is the internal vSmartMOM coordinate used in the file
names for the edge-on grid. The symbol $\alpha$ should be read as the
star-planet-observer phase angle; it does not imply that a real observed
system must be edge-on. For one epoch, the general state is
\begin{equation}
  \mathbf{z}=(u,\rhosurf,\alpha), \qquad u=\ln\taufour .
  \label{eq:app_fullstate}
\end{equation}
For $N$ epochs with unknown phases this becomes
\begin{equation}
  \mathbf{z}_N=(u,\rhosurf,\alpha_1,\ldots,\alpha_N).
  \label{eq:app_fullstateN}
\end{equation}
When phase is known, $\alpha$ is fixed externally and the inferred
{atmosphere-surface} state is
\begin{equation}
  \statevec=(u,\rhosurf), \qquad u=\ln\taufour .
  \label{eq:app_state}
\end{equation}

For each {atmosphere-surface} state and phase, the forward model returns the disk-integrated
quantities $I_{360}$, $I_{400}$, $Q_{400}$, $U_{400}$,
$\DOLP_{360}$, and $\DOLP_{400}$. The intensive observables are
\begin{align}
  \slopeI(\statevec,\alpha)
    &= \frac{I_{360}(\statevec,\alpha)
      - I_{400}(\statevec,\alpha)}
      {I_{400}(\statevec,\alpha)}, \\
  \DOLP_{400}(\statevec,\alpha)
    &= \frac{\sqrt{Q_{400}^2(\statevec,\alpha)
      + U_{400}^2(\statevec,\alpha)}}
      {I_{400}(\statevec,\alpha)}, \\
  \slopeD(\statevec,\alpha)
    &= \frac{\DOLP_{360}(\statevec,\alpha)
      - \DOLP_{400}(\statevec,\alpha)}
      {\DOLP_{400}(\statevec,\alpha)} .
  \label{eq:app_observables}
\end{align}
We collect them as
\begin{equation}
  \obsvec(\statevec,\alpha)
  =(\slopeI,\DOLP_{400},\slopeD)^{\top}.
  \label{eq:app_obsvec}
\end{equation}

\subsection{Local area metric}
\label{app:conditioning}

For known phase, the observable map is restricted to the
two-dimensional state $\statevec=(u,\rhosurf)^{\top}$. Its local
sensitivity is
\begin{equation}
  \jac(\statevec;\alpha)
  = \frac{\partial \obsvec}{\partial \statevec}.
  \label{eq:app_jacobian}
\end{equation}
The corresponding Gram matrix is
\begin{equation}
  G(\statevec;\alpha)
  =\jac(\statevec;\alpha)^{\top}\jac(\statevec;\alpha),
  \label{eq:app_gram}
\end{equation}
and the local area metric is
\begin{equation}
  \mathcal{A}(\statevec;\alpha)
  =\sqrt{\det G(\statevec;\alpha)} .
  \label{eq:app_areametric}
\end{equation}
This is the infinitesimal area in observable space corresponding to a
unit area in $(u,\rhosurf)$. For multiple known phases, the Gram
matrices add,
\begin{equation}
  G_{\mathrm{stack}}(\statevec)
  =
  \sum_i
  \jac(\statevec;\alpha_i)^{\top}
  \jac(\statevec;\alpha_i),
  \label{eq:app_gstack}
\end{equation}
so the plotted six-phase metric in Figure~\ref{fig:unique_snr} is
$\mathcal{A}_{6\alpha}=\sqrt{\det G_{\mathrm{stack}}}$ for
$\alpha=(150^\circ,120^\circ,90^\circ,60^\circ,30^\circ,5^\circ)$.
Distinct phases improve the conditioning only when their sensitivity
directions are not redundant.

The phase-dependent disk reflectivity used for the radius consequence is
\begin{equation}
  \Lambda(\statevec,\alpha)
  = \frac{I_{400}(\statevec,\alpha)}{\Idisk},
  \qquad
  \Idisk=\frac{3}{2}I_{400}(\taufour=0,\rhosurf=1,\alpha=0^\circ).
  \label{eq:app_lambda}
\end{equation}
The full-phase point $\alpha=0^\circ$ is used for this normalization,
but is excluded from the unknown-phase trial set because {it is not a
direct-imaging geometric configuration.}

For a stated SNR referenced to $I_{400}$, the propagated uncertainties
used to whiten the intensive observables are
\begin{align}
  \sigma(\slopeI)
     &= \frac{\sqrt{2}}{\mathrm{SNR}}
       \frac{I_{360}}{I_{400}}, \\
  \sigma(\DOLP_{\lambda})
    &= \frac{\sqrt{1+\DOLP_{\lambda}^2}}{\mathrm{SNR}}, \\
  \sigma(\slopeD)
    &= \frac{1}{|\DOLP_{400}|}
       \left[
       \sigma(\DOLP_{360})^2
       +(1+\slopeD)^2\sigma(\DOLP_{400})^2
       \right]^{1/2}.
  \label{eq:app_noise}
\end{align}
For a true state $a$ and trial state $b$, the symmetric whitening
scale for observable component $j$ is
\begin{equation}
  \bar{\sigma}_j(a,b)
  = \frac{1}{2}\,[\sigma_j(a)+\sigma_j(b)] .
  \label{eq:app_sigbar}
\end{equation}
Because $\slopeI$ becomes numerically unreliable when
$I_{360}$ and $I_{400}$ differ only at the disk-integration floor, we
include the intensity-color component in the distance metric only when
the spectral difference is above this floor for both the true and trial
states:
\begin{equation}
  |I_{360}(\statevec,\alpha)-I_{400}(\statevec,\alpha)|
  \ge \epsilon_{\mathrm{floor}},
  \qquad \epsilon_{\mathrm{floor}}=3\times10^{-5}.
  \label{eq:app_floor}
\end{equation}
The present implementation applies this floor only to $\slopeI$. An
analogous floor for $\slopeD$ would be required in regions where either
$\DOLP_{400}$ or the 360-400 nm DOLP difference approaches the
disk-integration floor.

\subsection{Single-epoch known and unknown phase}
\label{app:single_phase}

For a single observation at true phase $\alpha_{\mathrm{true}}$, the known-phase
trial set fixes the trial phase to the true phase. The
whitened distance is
\begin{equation}
  d_{\mathrm{known}}^2(a,b)
  = \sum_{j\in{\mathrm{kept}}}
    \left[
    \frac{
    m_j(\statevec_b,\alpha_{\mathrm{true}})
    -m_j(\statevec_a,\alpha_{\mathrm{true}})}
    {\bar{\sigma}_j(a,b)}
    \right]^2 .
  \label{eq:app_dknown}
\end{equation}
The posterior weight includes the state-cell volume,
\begin{equation}
  w_b(a) \propto
  \exp[-d_{\mathrm{known}}^2(a,b)/2]\,\Delta u_b\,\Delta\rho_b ,
  \qquad
  W_b(a)=\frac{w_b(a)}{\sum_c w_c(a)} .
  \label{eq:app_wknown}
\end{equation}

In the unknown-phase bound, the trial phase is also varied
over the sampled phase grid. The distance becomes
\begin{equation}
  d_{\mathrm{unknown}}^2(a,b)
  = \sum_{j\in{\mathrm{kept}}}
    \left[
    \frac{
    m_j(\statevec_b,\alpha_b)
    -m_j(\statevec_a,\alpha_{\mathrm{true}})}
    {\bar{\sigma}_j(a,b)}
    \right]^2 ,
  \label{eq:app_dunknown}
\end{equation}
and the posterior weight is
\begin{equation}
  w_b(a) \propto
  \exp[-d_{\mathrm{unknown}}^2(a,b)/2]\,
  \Delta u_b\,\Delta\rho_b\,\Delta\alpha_b ,
  \qquad
  W_b(a)=\frac{w_b(a)}{\sum_c w_c(a)} .
  \label{eq:app_wunknown}
\end{equation}

For a single epoch, the reflected flux satisfies
\begin{equation}
  F_{\mathrm{obs}} \propto \Rp^2\Lambda(\statevec,\alpha).
  \label{eq:app_flux_single}
\end{equation}
For the true state $a$ and trial state $b$, matching the same
measured flux gives
\begin{equation}
  {\Rp}_b^2\Lambda_b={\Rp}_a^2\Lambda_a ,
  \qquad
  \frac{{\Rp}_b}{{\Rp}_a}
  =\left(\frac{\Lambda_a}{\Lambda_b}\right)^{1/2}.
  \label{eq:app_radius_exact_single}
\end{equation}
For continuity with the albedo-degeneracy analysis in the main text, the
linearized consequence is reported as
\begin{equation}
  \frac{\sigma(\Rp)}{\Rp}
  \simeq \frac{1}{2}\frac{\sigma_\Lambda}{\Lambda_a},
  \label{eq:app_radius_linear}
\end{equation}
where
\begin{align}
  \bar{\Lambda}(a) &= \sum_b W_b(a)\Lambda_b, \\
  \sigma_\Lambda^2(a)
    &= \sum_b W_b(a)[\Lambda_b-\bar{\Lambda}(a)]^2 .
  \label{eq:app_lambdavar}
\end{align}

\subsection{Multi-epoch unknown-phase calculation}
\label{app:multi_phase}

For $N$ epochs, let the true phase tuple be
\begin{equation}
  A_{\mathrm{true}}
  =(\alpha_{\mathrm{true},1},\ldots,\alpha_{\mathrm{true},N}).
  \label{eq:app_Atrue}
\end{equation}
Known-phase trial states use $A_b=A_{\mathrm{true}}$. Unknown-phase trial states use an
ordered tuple from the sampled phase grid,
\begin{equation}
  A_b=(\alpha_{b,1},\ldots,\alpha_{b,N}).
  \label{eq:app_Ab}
\end{equation}
The stacked intensive vector is
\begin{equation}
  M_{\mathrm{int}}(\statevec,A)
  =
  [\obsvec(\statevec,\alpha_1),\ldots,
   \obsvec(\statevec,\alpha_N)] .
  \label{eq:app_Mint}
\end{equation}
Let $\alpha_{\rm ref}$ denote the reference phase, chosen here as the
brightest epoch in the multi-epoch set,
\begin{equation}
  \alpha_{\rm ref}
  =
  \alpha_{k_*}, \qquad
  k_*=\arg(\max_k (I_{400}(\statevec,\alpha_k))).
  \label{eq:app_alpha_ref}
\end{equation}
The normalized lightcurve-shape vector is
\begin{equation}
  \ell_i(\statevec,A)
  =
  \frac{I_{400}(\statevec,\alpha_i)}
       {I_{400}(\statevec,\alpha_{\rm ref})},
  \qquad i=1,\ldots,N .
  \label{eq:app_shape}
\end{equation}
Equivalently, after ordering the phase tuple so that
$\alpha_1=\alpha_{\rm ref}$, the shape vector is
$\ell_i=I_{400}(\alpha_i)/I_{400}(\alpha_1)$.
The full multi-epoch measurement vector is
\begin{equation}
  \mathbf{y}(\statevec,A)
  = [M_{\mathrm{int}}(\statevec,A),\ell(\statevec,A)] .
  \label{eq:app_y}
\end{equation}
The calculation uses the independent ratio approximation
\begin{equation}
  \sigma(\ell_i)=\frac{\sqrt{2}\,\ell_i}{\mathrm{SNR}},
  \label{eq:app_shape_noise}
\end{equation}
with the same symmetric whitening prescription as
Equation~\ref{eq:app_sigbar}.

The multi-epoch distance is
\begin{align}
  d_{\mathrm{multi}}^2(a,b)
  &=
  \sum_{k=1}^{N}
  \sum_{j\in{\mathrm{kept}}_k}
    \left[
    \frac{
    m_j(\statevec_b,\alpha_{b,k})
    -m_j(\statevec_a,\alpha_{\mathrm{true},k})}
    {\bar{\sigma}_{j,k}(a,b)}
    \right]^2 \nonumber \\
  &\quad +
  \sum_{k=1}^{N}
    \left[
    \frac{
    \ell_k(\statevec_b,A_b)
    -\ell_k(\statevec_a,A_{\mathrm{true}})}
    {\bar{\sigma}_{\ell,k}(a,b)}
    \right]^2 .
  \label{eq:app_dmulti}
\end{align}
The corresponding weights are
\begin{align}
  w_{b,\mathrm{known}}(a)
  &\propto
  \exp[-d_{\mathrm{multi}}^2(a,b)/2]\,
  \Delta u_b\,\Delta\rho_b, \\
  w_{b,\mathrm{unknown}}(a)
  &\propto
  \exp[-d_{\mathrm{multi}}^2(a,b)/2]\,
  \Delta u_b\,\Delta\rho_b
  \prod_{k=1}^{N}\Delta\alpha_{b,k},
  \label{eq:app_wmulti}
\end{align}
followed in each case by the normalization
$W_b(a)=w_b(a)/\sum_c w_c(a)$.

For the radius consequence, define the true-state and trial-state reflectivity
vectors
\begin{align}
  \boldsymbol{\Lambda}_a
  &=
  [\Lambda(\statevec_a,\alpha_{\mathrm{true},1}),\ldots,
   \Lambda(\statevec_a,\alpha_{\mathrm{true},N})], \\
  \boldsymbol{\Lambda}_b
  &=
  [\Lambda(\statevec_b,\alpha_{b,1}),\ldots,
   \Lambda(\statevec_b,\alpha_{b,N})].
  \label{eq:app_lambdavectors}
\end{align}
The observed flux vector is proportional to
${\Rp}_a^2\boldsymbol{\Lambda}_a$. A trial state uses a fitted scale
$s_b=({\Rp}_b/{\Rp}_a)^2$, chosen by least squares:
\begin{equation}
  s_b = \arg\min_s
  \sum_{i=1}^{N}
  (s\Lambda_{b,i}-\Lambda_{a,i})^2 .
  \label{eq:app_smin}
\end{equation}
Differentiating with respect to $s$ gives
\begin{equation}
  0 = 2\sum_i \Lambda_{b,i}(s_b\Lambda_{b,i}-\Lambda_{a,i}),
  \label{eq:app_sderiv}
\end{equation}
and therefore
\begin{equation}
  s_b =
  \frac{\boldsymbol{\Lambda}_b\cdot\boldsymbol{\Lambda}_a}
       {\boldsymbol{\Lambda}_b\cdot\boldsymbol{\Lambda}_b},
  \qquad
  r_b\equiv\frac{{\Rp}_b}{{\Rp}_a}
  =
  \left[
  \frac{\boldsymbol{\Lambda}_b\cdot\boldsymbol{\Lambda}_a}
       {\boldsymbol{\Lambda}_b\cdot\boldsymbol{\Lambda}_b}
  \right]^{1/2}.
  \label{eq:app_rmulti}
\end{equation}
The plotted multi-epoch radius consequence is the posterior width of
$r_b$,
\begin{align}
  \bar{r}(a) &= \sum_b W_b(a)r_b, \\
  \sigma_r^2(a) &= \sum_b W_b(a)[r_b-\bar{r}(a)]^2, \\
  100\,\frac{\sigma(\Rp)}{\Rp} &= 100\,\sigma_r(a).
  \label{eq:app_sigmar}
\end{align}
The diagnostic RMSE about the true radius is
\begin{equation}
  {\mathrm{RMSE}}_r(a)
  =
  \left[
  \sum_b W_b(a)(r_b-1)^2
  \right]^{1/2}.
  \label{eq:app_rmse}
\end{equation}

For the unknown-phase case, the posterior phase spread at epoch $k$ is
\begin{align}
  \bar{\alpha}_k(a) &= \sum_b W_b(a)\alpha_{b,k}, \\
  \sigma_{\alpha,k}^2(a)
    &= \sum_b W_b(a)[\alpha_{b,k}-\bar{\alpha}_k(a)]^2 ,
  \label{eq:app_phase_spread}
\end{align}
and the reported diagnostic phase spread is the epoch average
\begin{equation}
  \sigma_{\alpha,\mathrm{reported}}(a)
  = \frac{1}{N}\sum_{k=1}^{N}\sigma_{\alpha,k}(a).
  \label{eq:app_phase_reported}
\end{equation}
Finally, the known-phase and unknown-phase cases are limiting
forms of a more general phase-prior calculation. A prior $p(A_b)$ would
replace the uniform phase-volume factor by
\begin{equation}
  \prod_{k=1}^{N}\Delta\alpha_{b,k}
  \quad\longrightarrow\quad
  p(A_b)\prod_{k=1}^{N}\Delta\alpha_{b,k}.
  \label{eq:app_phase_prior}
\end{equation}

\section{Continuous inversion used for the HWO radius consequences}
\label{app:hwo_inversion}

The {HWO-simulation} radius consequences in
Figure~\ref{fig:hwo_radius_consequences} are computed with a continuous
lookup-table inversion, {rather than by summing over the discrete trial
states used in Appendix~\ref{app:phase_unknown}.} For $N=6$ epochs, the
{inference} vector is
\begin{equation}
  \statevec
  =
  (u,\rhosurf,\alpha_1,\ldots,\alpha_N)^{\top},
  \qquad
  u=\ln\taufour .
  \label{eq:app_hwo_inv_state}
\end{equation}
For a given {HWO case}, the allowed state domain is
\begin{equation}
  \mathcal{D}_{\mathrm{HWO}}
  =
  \left\{
  \statevec:
  u_{\min}\le u\le u_{\max},\;
  \rho_{\min}\le\rhosurf\le\rho_{\max},\;
  \alpha_{\min}\le\alpha_i\le\alpha_{\max}
  \right\},
  \label{eq:app_hwo_inv_domain}
\end{equation}
where the phase bounds are the IWA- and exposure-limited interval for
the corresponding target distance. The forward model is the
Horak/\vSmartMOM\ table evaluated by cubic interpolation. For any
tabulated quantity $q$, we write this interpolated value as
\begin{equation}
  \widetilde{q}(u,\rhosurf,\alpha)
  =
  \mathcal{I}_{\mathrm{cub}}[q](u,\rhosurf,\alpha).
  \label{eq:app_hwo_inv_interp}
\end{equation}

The normalized 400 nm flux, or lightcurve-shape component, divides out
one common brightness scale:
\begin{equation}
  {\shapeI}_i(\statevec)
  =
  \frac{\widetilde{I}_{400}(u,\rhosurf,\alpha_i)}
       {I_{\mathrm{ref}}(\statevec)},
  \qquad i=1,\ldots,N .
  \label{eq:app_hwo_inv_shape}
\end{equation}
Here $I_{\mathrm{ref}}$ is a fixed reference intensity from the same
modeled phase set. One component of ${\shapeI}_i$ is therefore unity by
construction and carries no independent shape information.

At each epoch, the full spectropolarimetric observable vector is
\begin{equation}
  \mathbf{m}_i^{\mathrm{full}}(\statevec)
  =
  \left[
  {\shapeI}_i(\statevec),\;
  \slopeI(u,\rhosurf,\alpha_i),\;
  \DOLP_{400}(u,\rhosurf,\alpha_i),\;
  \slopeD(u,\rhosurf,\alpha_i)
  \right]^{\top},
  \label{eq:app_hwo_inv_mfull}
\end{equation}
with the spectral and polarization components defined as in
Equation~\ref{eq:app_observables}, but evaluated using the interpolated
table. The no-polarimetry case retains only the intensity-color and
normalized-flux components,
\begin{equation}
  \mathbf{m}_i^{\mathrm{nopol}}(\statevec)
  =
  \left[
  {\shapeI}_i(\statevec),\;
  \slopeI(u,\rhosurf,\alpha_i)
  \right]^{\top}.
  \label{eq:app_hwo_inv_mnopol}
\end{equation}
Thus the modeled measurement vector for observable mode
$o\in\{\mathrm{full},\mathrm{nopol}\}$ is
\begin{equation}
  \mathbf{y}_{o}(\statevec)
  =
  \mathbf{P}_{\shapeI}
  \left[
  \mathbf{m}_1^{o},\ldots,
  \mathbf{m}_N^{o}
  \right]^{\top}.
  \label{eq:app_hwo_inv_y}
\end{equation}

The synthetic observation is generated at the retained phase set
$\statevec_\ast$,
\begin{equation}
  \mathbf{y}_{\ast,o}
  =
  \mathbf{y}_{o}(\statevec_\ast),
  \label{eq:app_hwo_inv_ystar}
\end{equation}
and the {inversion} uses the full measurement covariance
$\mathbf{C}_{o}$ for the chosen SNR. The objective function is
\begin{equation}
  \chi^2_{o}(\statevec)
  =
  \left[
  \mathbf{y}_{o}(\statevec)
  -\mathbf{y}_{\ast,o}
  \right]^{\top}
  \mathbf{C}_{o}^{-1}
  \left[
  \mathbf{y}_{o}(\statevec)
  -\mathbf{y}_{\ast,o}
  \right] .
  \label{eq:app_hwo_inv_chi2}
\end{equation}
The reported best fit is
\begin{equation}
  \widehat{\statevec}
  =
  \arg\min_{\statevec\in\mathcal{D}_{\mathrm{HWO}}}
  \chi^2_{o}(\statevec).
  \label{eq:app_hwo_inv_xhat}
\end{equation}
Numerically, the minimization is carried out as a multistart search in
the transformed bounded space; the lowest-$\chi^2$ converged result is
retained.

The local covariance is then computed from the linearized map at the
best-fitting state. Let
\begin{equation}
  \mathbf{J}_{o}
  =
  \left.
  \frac{\partial \mathbf{y}_{o}}
       {\partial \statevec}
  \right|_{\widehat{\statevec}} .
  \label{eq:app_hwo_inv_jac}
\end{equation}
The Gaussian approximation to the posterior is
\begin{equation}
  \chi^2_{o}(\widehat{\statevec}+\delta\statevec)
  \simeq
  \chi^2_{o}(\widehat{\statevec})
  +
  \delta\statevec^{\top}
  \mathbf{H}_{o}
  \delta\statevec,
  \qquad
  \mathbf{H}_{o}
  =
  \mathbf{J}_{o}^{\top}
  \mathbf{C}_{o}^{-1}
  \mathbf{J}_{o},
  \label{eq:app_hwo_inv_hessian}
\end{equation}
so that
\begin{equation}
  \boldsymbol{\Sigma}_{\statevec}
  =
  \mathbf{H}_{o}^{-1}.
  \label{eq:app_hwo_inv_cov}
\end{equation}

{The radius is not fitted in the intensive-observable inversion. It is
introduced only after the {atmosphere-surface-phase} covariance has been
estimated.} Define the six-epoch disk-brightness vector
\begin{equation}
  \boldsymbol{\Lambda}(\statevec)
  =
  \left[
  \Lambda(u,\rhosurf,\alpha_1),\ldots,
  \Lambda(u,\rhosurf,\alpha_N)
  \right]^{\top},
  \label{eq:app_hwo_inv_lambdavec}
\end{equation}
with $\Lambda$ defined by Equation~\ref{eq:app_lambda}. A perturbation
$\delta\statevec$ changes the least-squares radius scale required to
match the same absolute 400 nm fluxes. Linearizing
Equation~\ref{eq:app_rmulti} gives
\begin{equation}
  \delta\ln\Rp
  =
  \mathbf{g}_{R}^{\top}\delta\statevec,
  \qquad
  \mathbf{g}_{R}^{\top}
  =
  -\frac{1}{2}
  \frac{
  \boldsymbol{\Lambda}^{\top}
  \left(\partial\boldsymbol{\Lambda}/\partial\statevec\right)}
  {\boldsymbol{\Lambda}^{\top}\boldsymbol{\Lambda}}
  \bigg|_{\widehat{\statevec}} .
  \label{eq:app_hwo_inv_gradR}
\end{equation}
The plotted local-covariance radius consequence is therefore
\begin{equation}
  100\,\frac{\sigma(\Rp)}{\Rp}
  =
  100
  \left(
  \mathbf{g}_{R}^{\top}
  \boldsymbol{\Sigma}_{\statevec}
  \mathbf{g}_{R}
  \right)^{1/2}.
  \label{eq:app_hwo_inv_radius}
\end{equation}
Because the synthetic observations in this diagnostic are {evaluated at their exact model values, with the astrophysical noise budget affecting only the covariance,}
the best-fitting states recover $\statevec_\ast$ to numerical
precision. The plotted width is the local consequence of the adopted
SNR and observable set, not the error of a noisy realization.

\section{HWO UV-A exposure-time formalism}
\label{app:hwo_exptime}

This appendix documents the exposure-time calculation used for the
{six-phase HWO simulations} in
Section~\ref{sec:hwo_example}. For each trial phase angle,
exposure times are computed independently in two 10 nm UV-A bands,
355-365 nm and 395-405 nm, represented by the 360 nm and 400 nm
Horak/\vSmartMOM\ reflectivities. The required exposure time for that
phase is the slower of the two bands. This ensures that both the
intensity and polarization-slope observables can be measured at the
target SNR.

Let $\Phi_{\star,\lambda}(\lambda)$ be the stellar photon spectral flux
density at the observer, in photons m$^{-2}$ s$^{-1}$ nm$^{-1}$. For
the synthetic solar twins at distance $d$, this is computed from the
ASTM E490 AM0 solar spectrum as
\begin{equation}
  \Phi_{\star,\lambda}(\lambda;d)
  =
  F_{\mathrm{E490}}(\lambda)
  \frac{\lambda}{hc}
  \left(\frac{\mathrm{AU}}{d}\right)^2 ,
  \label{eq:app_hwo_phistar}
\end{equation}
where $F_{\mathrm{E490}}(\lambda)$ is the solar irradiance at 1 AU in
W m$^{-2}$ nm$^{-1}$. The band-integrated photon flux is
\begin{equation}
  \Phi_{\star,\mathrm{band}}(d)
  =
  \int_{\mathrm{band}}
  \Phi_{\star,\lambda}(\lambda;d)\,d\lambda .
  \label{eq:app_hwo_bandflux}
\end{equation}
For real host stars, the implementation applies a simple blackbody color
correction using the stellar effective temperature and luminosity. For
the constructed solar twins used here, the correction is unity.

The detected stellar count rate before coronagraphic suppression is
\begin{equation}
  N_s
  =
  \Phi_{\star,\mathrm{band}}(d)
  A_{\mathrm{tel}}T,
  \qquad
  A_{\mathrm{tel}}=\frac{\pi D_{\mathrm{tel}}^2}{4},
  \label{eq:app_hwo_ns}
\end{equation}
with $D_{\mathrm{tel}}=8$ m and $T=0.1$ in the numerical examples. The
phase-dependent planet-star contrast is
\begin{equation}
  \epsilon(\alpha)
  =
  \Lambda(\taufour,\rhosurf,\alpha)
  \left(\frac{\Rp}{a}\right)^2 ,
  \label{eq:app_hwo_contrast}
\end{equation}
where $\Lambda=I_{\mathrm{band}}/\Idisk$ is obtained from the
Horak/\vSmartMOM\ disk-integrated forward model. The detected planet
count rate is
\begin{equation}
  CR_p
  =
  \Upsilon\,\epsilon(\alpha)\,N_s ,
  \label{eq:app_hwo_crp}
\end{equation}
with $\Upsilon=0.69$ inside the Stark photometric aperture
$X=0.7\lambda/D_{\mathrm{tel}}$.

The total background count rate is
\begin{equation}
  CR_b
  =
  CR_{\mathrm{star\,leak}}
  +CR_{\mathrm{zodi}}
  +CR_{\mathrm{exozodi}} .
  \label{eq:app_hwo_crb}
\end{equation}
The residual stellar leakage term is
\begin{equation}
  CR_{\mathrm{star\,leak}}
  =
  \Upsilon\,RC\,N_s,
  \qquad
  RC=10^{-10}.
  \label{eq:app_hwo_leak}
\end{equation}
The photometric aperture solid angle is
\begin{equation}
  \Omega_{\mathrm{ap}}
  =
  \pi\left(\frac{X\lambda_c}{D_{\mathrm{tel}}}\right)^2 ,
  \label{eq:app_hwo_omega}
\end{equation}
where $\lambda_c$ is the band center. The local-zodi and exozodi count
rates are
\begin{align}
  CR_{\mathrm{zodi}}
  &=
  B_{\mathrm{zodi,band}}\,
  \Omega_{\mathrm{ap}}A_{\mathrm{tel}}T, \\
  CR_{\mathrm{exozodi}}
  &=
  n_{\mathrm{exozodi}}\,
  B_{\mathrm{exozodi,band}}\,
  \Omega_{\mathrm{ap}}A_{\mathrm{tel}}T .
  \label{eq:app_hwo_dust}
\end{align}
The UV-A surface brightnesses are color-corrected from the V-band Stark
surface-brightness anchors,
\begin{align}
  B_{\mathrm{zodi,band}}
  &=
  \frac{1}{\theta_{\mathrm{arcsec}}^2}
  \int_{\mathrm{band}}
  F_{0,V}10^{-0.4z}
  \frac{\Phi_\odot(\lambda)}{\Phi_\odot(550~\mathrm{nm})}
  \,d\lambda, \\
  B_{\mathrm{exozodi,band}}
  &=
  \frac{1}{\theta_{\mathrm{arcsec}}^2}
  \int_{\mathrm{band}}
  F_{0,V}10^{-0.4x}
  \frac{\Phi_\star(\lambda)}{\Phi_\star(550~\mathrm{nm})}
  \,d\lambda,
  \label{eq:app_hwo_dust_brightness}
\end{align}
with $F_{0,V}=9500\times10^{13}$ photons m$^{-3}$ s$^{-1}$,
$z=23$ mag arcsec$^{-2}$, $x=22$ mag arcsec$^{-2}$,
$n_{\mathrm{exozodi}}=3$, and
$\theta_{\mathrm{arcsec}}=\pi/(180\times3600)$.

The photon-limited Stark exposure time is
\begin{equation}
  t_{\mathrm{Stark}}
  =
  \mathrm{SNR}^2
  \frac{CR_p+2CR_b}{CR_p^2}.
  \label{eq:app_hwo_stark}
\end{equation}
The factor of two multiplying $CR_b$ follows the conservative
background-subtraction convention used by \citet{Stark2014}; the planet
shot-noise term is included once. The corresponding Mennesson-style
implementation is
\begin{equation}
  t_{\mathrm{Mennesson}}
  =
  \mathrm{SNR}^2
  \frac{CR_p+2CR_b}
  {CR_p^2-\mathrm{SNR}^2
  (\Upsilon\sigma_{\Delta C}N_s)^2}.
  \label{eq:app_hwo_mennesson}
\end{equation}
For all {HWO examples} in this manuscript bundle,
$\sigma_{\Delta C}=0$, so $t_{\mathrm{Mennesson}}=t_{\mathrm{Stark}}$.

\section*{Acknowledgements}
 The work presented herein is supported by the National Aeronautics and Space Administration under a contract issued through the Astrophysics Division of the Science Mission Directorate (80NM0018D0004). This research was carried out at the Jet Propulsion Laboratory, California Institute of Technology, under contract with NASA. Copyright 2026.  All rights reserved.

\bibliography{refs_assimilated}
\bibliographystyle{aasjournal}

\end{document}